\documentstyle[preprint,aps]{revtex}
\newcommand{\be}{\begin{equation}}
\newcommand{\ee}{\end{equation}}
\newcommand{\bear}{\begin{eqnarray}}
\newcommand{\ear}{\end{eqnarray}}
\newcommand{\ba}{\begin{eqnarray*}}
\newcommand{\ea}{\end{eqnarray*}}

\newcommand{\D}{\cal D}
\newcommand{\N}{\cal N}
\newcommand{\B}{\cal B}

\newcommand{\Z}{\mbox{\Large${\cal Z}$}}

\newcommand{\opF}{\mbox{\boldmath${\cal F}$}}
\newcommand{\A}{\cal A}

\begin{document}

\title{Structural aspects of the fermion-boson mapping in 
two-dimensional gauge and anomalous gauge
theories with massive fermions}

\author{L. V. Belvedere, A. de Sousa  Dutra$^\ast$, C. P. Natividade and A. F. de 
Queiroz 
\\
%}\address{
\small{\it Instituto de F\'{\i}sica - Universidade Federal Fluminense}\\
\small {\it Av. Litor\^anea, S/N, Boa Viagem, Niter\'oi}\\
\small{\it CEP.24210-340, Rio de Janeiro - Brasil}\\
$\ast$ \small{\it DFQ - UNESP - Campos de Guaratinguet\'a}\\
\small{\it Guaratinguet\'a, 12500, SP - Brasil}}

\date{\today}

\maketitle
%%%%%%%%%%%%%%%%%%%%%%%%%%%%%%%%%%%%%%%%%%%%%%%%%%%%%%%%%%%%%%%
\begin{abstract}

Using a synthesis of the functional integral and operator approaches we 
discuss the  fermion-boson
mapping and the role played by the Bose field algebra in the Hilbert space
of two-dimensional gauge and anomalous gauge
field theories with massive fermions. In the $QED_2$ with quartic 
self-interaction 
among massive fermions, the use of an auxiliary
vector field introduces a redundant Bose field algebra that should not be
considered as an element of the intrinsic algebraic structure defining
the model. In the anomalous
chiral $QED_2$ with massive fermions the effect of the chiral 
anomaly leads to the appearance in the mass operator of a spurious
Bose field combination. This phase factor carries no fermion selection rule  
and the expected absence of $\theta$-vacuum in the anomalous model 
is displayed from the operator solution. Even in the anomalous
model with massive Fermi fields, the introduction of
the Wess-Zumino field replicates the
theory, changing neither its algebraic content nor its physical content.
\end{abstract}

\newpage

%%%%%%%%%%%%%%%%%%%%%%%%%%%%%%%%%%%%%%%%%%%%%%%%%%%%%%%%%%%%%%%%%%%%%%
%%%%%%%%%%%%%%%%%%%%%%%%%%%%%%%%%%%%%%%%%%%%%%
\section{Introduction}
%%%%%%%%%%%%%%%%%%%%%%%%%%%%%%%%%%%%%%%%%%%%%%%%%%
\setcounter{equation}{0}

In the recent efforts towards the extension of the 
functional integral bosonization to $2 + 1$ 
dimensions \cite{Shapo,fradshapo}, use has 
been made of an interpolating field procedure that leads to a
``mapping'' of the partition function of the original theory into a partition 
function of Chern-Simons-type theories. Unfortunately, until recent 
investigations of the fermion-boson mappings  in $2 + 1$ dimensions were
limited to perturbative analysis and these mappings generally are 
established on the level of factorizable partition functions and  not
established on the Hilbert space of states.

The Thirring model in $2+1$ dimensions has been considered
in Refs.\cite{Shapo,fradshapo}. In the Abelian case, it has been shown 
that to lowest 
order in inverse fermion mass, the Thirring model partition function 
coincides
with that of the Maxwell-Chern-Simons theory. In the non-Abelian case, and 
in the low-energy
regime, it has been shown that the bosonized partition function of 
the $SU (N)$ massive Thirring model is
related to $SU (N)$ Yang-Mills-Chern-Simons gauge theory. In both cases, the
bosonization is performed using an auxiliary vector field and the 
correspondence
between the models are established on the level of the partition functions. 

At the present state of the research, and due to a large number of papers on
the subject, it seems to be very instructive to make a 
investigation of the basic structural aspects involved in the
functional integral bosonization approach using the auxiliary vector
field. The bosonization of two-dimensional gauge and anomalous gauge models 
has been reexamined
quite recently in Refs.\cite{FIB,M,Bel,BRR}. In Ref.\cite{FIB} using the 
functional integral formulation  we reconstruct in the Hilbert space of 
states the Coleman's proof \cite{C} of the fermion-boson mapping 
between the massive Thirring and sine-Gordon theories. We show that the 
use of an auxiliary vector field enlarges
the Hilbert space by the introduction of an external Bose field algebra that
should not be considered as an element of the intrinsic algebraic
structure defining the model. The factorization of the partition function
generally lead to incorrect conclusion concerning the physical content of the 
model \cite{FIB,Bel,BRR}. 

The general procedure to establish the fermion-boson mappings is performed
by using one of the quantum field 
theory approaches, the 
operator approach or the functional integral approach. However, many structural
aspects of the models are not  easily accessible or not visible at all in 
one of these approaches. In our presentation,  which 
interpolates between
the functional integral and operator approaches,  close attention 
is paid to mantaining complete control on the Hilbert space structure needed 
for the representation of the intrinsic field algebra, whose Wightman 
functions define the model.  The subject of this paper is to analyze 
non-perturbative aspects and the Hilbert space structure of 
two-dimensional gauge and anomalous gauge models with
a procedure which interpolates between the operator and functional integral 
formalisms. In this way,  the conclusions about 
the structural aspects of the models emerges as a synthesis 
of the analysis performed using the two quantum field theory approaches.

In order to obtain insight into the role played by the 
auxiliary vector field in the Hilbert space of the bosonized theory, in 
the first part of the present work we shall consider the functional
integral bosonization of the massive $QED_2$ with quartic fermion 
self-interaction (massive Schwinger-Thirring model).  Using the functional 
integral bosonization we obtain the operator solution and the 
fermion-boson mapping
is established in the Hilbert space of states. We show that the use of the 
auxiliary vector field
to reduce the Thirring interaction to a quadratic form, introduces a 
redundant Bose field algebra that should not be
considered as an element of the intrinsic algebraic structure defining
the model. The infinitely delocalized 
states generated from the Bose field combination with zero scale dimension do
not belong to the Hilbert space of the model. This streamlines
the discussion present in Refs.\cite{FIB,Shaposnik,Naon}.  
  
In the second part of the paper we consider the functional 
integral bosonization
of the anomalous chiral $QED_2$ with massive fermions. In Refs.\cite{Bel,BRR}
some structural aspects of the bosonization of 
anomalous two-dimensional models with massless fermions has been 
considered. The effective
bosonized partition function of the massless 
chiral $QED_2$ (massless chiral Schwinger model) factorizes into the
partition functions of a free
massive vector field, a massless free field carrying the free fermion selection
rules and two ``decoupled'' massless free fields quantized with 
opposite metric. In the model with massless fermions, the extraction of 
these ``decoupled'' free and massless 
Bose fields from the operator solution \cite{Carena}, by performing an improper
factorization of 
the Hilbert space representing the intrinsic field algebra, leads to 
some misleading conclusions about basic structural properties of 
the model, such as the equivalence of the Schwinger model with the chiral model
defined for the regularization depending parameter $a = 2$, the violation of 
the asymptotic factorization property of the Wightman functions and 
the existence of $\theta$-vacuum. These
features cannot be regarded as being structural properties of the 
anomalous model
since they are dependent of the use of a redundant Bose field 
algebra, rather than
on the field algebra which defines the model.  The BRST constraints and the
Hilbert space structure of
the isomorphic gauge noninvariant and gauge invariant bosonized formulations of
chiral $QCD_2$ with massless fermions was 
analyzed in Ref. \cite{BRR}. The BRST subsidiary conditions are found
not to provide a sufficient criterium for defining physical 
states in the Hilbert space
and additional superselection rules must be taken into account. 

In order to probe the effect of the anomaly in the mass term for the Fermi 
field, we 
consider the anomalous chiral $QED_2$  with massive fermions. We show that 
the  ``decoupled''
fields quantized with opposite metric acquire a non-trivial dynamics and
are coupled by a sine-Gordon-like interaction. Nevertheless, their combination
is a massless free field which generates zero norm states from the vacuum. The 
effect of the chiral 
anomaly leads to the appearance of a spurious
Bose field combination in the mass operator. This phase factor signalize the
explicit breakdown of the chiral symmetry and its only effect in the operator
solution is to ensure the 
invariance of the field algebra under extended local gauge 
transformations, allowing the existence of two 
isomorphic representations of the intrinsic 
field algebra. The expected absence of $\theta$-vacuum in the anomalous 
model \cite{Bel} is displayed from
the operator solution. We show the isomorphism between the gauge-noinvariant 
and gauge invariant field algebras. The role played  by the 
Wess-Zumino (WZ) field in the anomalous chiral model with massive Fermi fields
is the same played in the corresponding massless model. The introduction of the
WZ field replicates the theory, changing neither its algebraic structure nor its
physical content. This also streamlines the presentation 
of Refs.\cite{Bel,Boy}.  In order to obtain some insight into the fermion-boson
mapping of the corresponding non-Abelian models, we shall consider
the bosonization method using the Abelian reduction of the Wess-Zumino-Witten 
theory \cite{BRR,WZW,PW}.

%%%%%%%%%%%%%%%%%%%%%%%%%%%%%%%%%%%%%%%%%%%%%%%%%%%%%%%%%%%%%%%%%%%%%%
\section{Massive $QED_2$ with quartic self-interaction}
%%%%%%%%%%%%%%%%%%%%%%%%%%%%%%%%%%%%%%%%%%%%%%%%%%%%%%%%%%%%%%%%%%%%%%
\setcounter{equation}{0}

The two-dimensional quantum 
electrodynamics with quartic self-interaction
among massive fermions (massive Schwinger-Thirring model) is defined by the
formal Lagrangian \cite{Dutra},

\be\label{lagr}
{\cal L} = - \frac{1}{4} {\cal F}^2_{\mu \nu} + 
\bar \psi (i \slash\!\!\!\partial - e\,
\slash\!\!\!\!{\cal A} - M_o ) \psi -  \frac{g^2}{2}\,J ^\mu\,J _\mu\,,
\ee

\noindent where $J ^\mu$ is the fermionic current  \footnote{Our conventions 
are: $ x^\pm = x^0 \pm x^1;\, \partial_\pm = \partial_0 \pm \partial_1;\,
 {\cal A}^\pm = {\cal A}^0 \pm {\cal A}^1$;
$$ g^{00} = 1 = - g^{11}; \epsilon^{01} = - \epsilon^{10} = 1;
\gamma^0 = \pmatrix{0 & 1 \cr 1 & 0},  \gamma^1 = 
\pmatrix{0 & 1 \cr - 1 & 0};  \gamma^5 = \gamma^0 \gamma^1; 
\gamma^\mu \gamma^5 = 
\epsilon^{\mu \nu} \gamma_\nu\,,\,\tilde \partial_\mu = \epsilon_{\mu \nu} \partial^\nu\,.$$
The scalar and pseudoscalar massless free fields are decomposed as $
\phi (x) = \phi (x^+) + \phi (x^-),\phi (x) = \phi (x^+) - \phi (x^-)\,,
$ such that $\partial_\mu \,\phi (x) = \epsilon_{\mu \nu}\,\partial ^\nu\,
\phi (x)$.}
, $J ^\mu = \bar \psi \gamma ^\mu \psi $,  and the field-strength tensor 
is ${\cal F}_{\nu \mu} \doteq 
\partial_\mu  {\cal A}_\nu - \partial_\nu {\cal A}_\mu$.

Within the operator formulation, the intrinsic local field algebra $\Im$ 
defining the model is generated from the set of local field 
operators $\{\bar \psi,\psi,{\A}_\mu \}$, such 
that, $ \Im = \Im\{\bar \psi,\psi,{\A}_\mu \}$. From the 
vacuum expectation values of products of 
fields $\Phi (x) \in \Im$, we obtain the general Wightman 
functions $W^{(n)}$ of the
theory

\be
W^{(n)} (x_1,\dots,x_n) = \langle \Omega \vert \Phi (x_1) \cdots \Phi (x_n) \vert
\Omega \rangle \,,
\ee

\noindent where $\Omega$ denotes the state vector of the vacuum. From the 
Wightman functions $W^{(n)}$ the Hilbert space  ${\cal H}$ can be constructed. The 
physical content of the model is given by the local gauge invariant field 
subalgebra  $\Im_{_{\!gi}} \subset \Im$, which is represented in the 
gauge invariant subspace ${\cal H}_{_{\!gi}} \subset {\cal H}$, and built
from the gauge invariant Wightman functions.

In the functional integral formulation of  quantum field theory, the 
Euclidian Green functions are given by the averages

\be
G^{(n)} (x_1,\cdots,x_n) = \frac{\int\,\Phi (x_1) \cdots \Phi (x_n)
d\,\mu\,(\Phi )\,}{\int\,d\,\mu\,(\Phi )}\,,
\ee

\noindent where $d\,\mu\,(\Phi )$ is the probability measure in terms of the 
Euclidian action $S (\Phi )$,

\be
d \mu (\Phi ) = e^{\,\textstyle\,-\,S [\Phi]} \,{\cal D} \Phi\,.
\ee

\noindent For Green functions satisfying the Osterwalder-Schrader 
axioms \cite{OS}, the reconstruction theorem guarantees that the 
functions $\{G^{(n)}\}$ uniquely determine real-time Wightman 
distributions by analytic continuation in the time variables.

In
what follows we shall consider two-dimensional gauge and anomalous gauge models
using
functional integral bosonization \footnote{We shall consider the functional 
integral in Minkowski space and analytic continuation
to Euclidian field theory is  understood  everywhere.} and making a 
connexion with
the operator formulation. In the present approach, which interpolates between
the functional integral and operator approaches,  close attention 
is paid to mantaining complete control on the Hilbert space structure 
needed for
the representation of the intrinsic field algebra generated by the set of 
fundamental
fields $\{\bar \psi,\psi,{\A}_\mu \}$ whose Wightman functions define the 
model.

%%%%%%%%%%%%%%%%%%%%%%%%%%%%%%%%%%%%%%%%%%%%%%%%%%
\subsection{Functional integral bosonization}
%%%%%%%%%%%%%%%%%%%%%%%%%%%%%%%%%%%%%%%%%%%%%%%%%%

To begin with, consider the generating functional in terms of
the external Grassmann valued 
sources $\upsilon$, $\bar \upsilon$, and the vector source $\omega_\mu$,

\be\label{gf}
{\Z}  [ \upsilon,\bar \upsilon,\omega_\mu ] = 
\langle\, e^{\,\textstyle i\,(\,\bar \psi\, \upsilon + \bar \upsilon \,\psi  + 
\A_\mu \,\omega^\mu\,)}\, \rangle_{_{\!\Omega}}\, = \,
 {\N}^{-1}\,
\int\,d \mu\,\,e^{\,\textstyle i\,\int d^2 z\,(\bar \psi \upsilon + \bar \upsilon \psi  + 
\A_\mu \omega^\mu)}\,,
\ee

\noindent where the functional integral  measure is given by

\be
d \mu\,=\,{\cal D} {\cal A}_\mu\,
{\cal D}\,\bar \psi\,{\cal D}\,\psi\,\,
e^{\,\textstyle{ i\,  S [\bar\psi,\psi,{\cal A}_\mu]}}\,,
\ee

\noindent and $ S [\bar \psi,\psi,{\cal A}_\mu] = 
\int\,d^2 x\,{\cal L} $, is the action corresponding to (\ref{lagr}).

Due to the presence of the quartic Fermi field interaction, the first step in
the standard procedure of the functional integral 
bosonization \cite{Shapo,FIB,Shaposnik,Naon} is 
performed with the help of an ``auxiliary'' vector 
field $b_\mu$. As stressed in Ref.\cite{FIB}, the 
introduction of the auxiliary vector field  defines an 
enlarged field algebra $\Im^\prime$, defined by $\Im^\prime = 
\Im^\prime \{\bar \psi,\psi,{\cal A}_\mu, b_\mu\}$, such that  $\Im \subset
\Im^\prime$. The field algebra $\Im^\prime$ is represented in the Hilbert space
${\cal H}^\prime = \Im^\prime \Omega \supset {\cal H} $. This is donne
by defining a new functional integral measure

\be\label{nm}
d \mu^\prime = d \mu\,{\cal D}\,b_\mu\,
e^{\,\textstyle i\,\int\,d^2 x\,\frac{1}{2}\, b^\mu b_\mu}\,.
\ee

\noindent Introducing the source term for the ``auxiliary'' vector field, the
Hilbert space ${\cal H}^\prime$ can be built from 
the ``interpolating'' generating functional 

$$
 \Z ^{\prime}  [ \upsilon,\bar{\upsilon},\omega_\mu,\rho_\mu ] =
\langle\, e^{\,\textstyle i\,\int d^2 z\,
(\,\bar \psi\, \upsilon + \bar \upsilon \,\psi  + 
{\cal A}_\mu \,\omega^\mu\, +\, b_\mu \rho^\mu\,)}\, \rangle^\prime_{_{\!\Omega}}\, =
$$

\be\label{mgf}
 =  {\N}^{-1}\,
\int\,d \mu^\prime \,
e^{\,\textstyle i\,\int\,d^2 x\,
(\bar \psi \upsilon + \bar \upsilon \psi  + 
{\cal A}_\mu \omega^\mu + b^\mu \rho_\mu )}
\,.
\ee

\noindent The source term for the field $b_\mu$ was 
included in 
order to control the effects of the auxiliary vector field on the 
bosonization procedure and the construction of the enlarged Hilbert 
space ${\cal H}^\prime$.   

In order to reduce the
action of the Thirring model to a quadratic form in the Fermi
field, the auxiliary vector field is shiffted 
by \cite{Shapo,fradshapo,FIB,Shaposnik}

\be\label{cv1}
b_\mu = B_\mu - g \bar \psi \gamma_\mu \psi\,,
\ee

\noindent in a  such way that

\be
\int\,{\D} b_\mu\,
e^{\, i\,\int\,d^2 x\,\frac{1}{2}\,\{\,b^\mu b_\mu\, -  \,g^2\,
J ^\mu\,J _\mu \}}   = \int {\cal D} 
B_\mu\, e^{\,i\,\int\,d^2x\,
 \{ \,\frac{1}{2}\,B_\mu\,B^\mu\,-\,g\,
J ^\mu\,B_{\mu}  \}}\,.
\ee

\noindent This leads to a effective 
action  in (\ref{mgf}), which is given in terms of the Lagrangian density

\be\label{l1}
{\cal L}^\prime_{_{eff.}} = -\,\frac{1}{4}\,{\cal F}_{\mu \nu}^2 \,+\,
 \bar \psi\,\slash\!\!\!\!D ({\cal A},B) \psi 
- m_o \bar \psi \,\psi \,+\,\frac{1}{2}\,B_\mu\,B^\mu\,,
\ee

\noindent where the covariant derivative is defined 
by $ \slash\!\!\!\!D ({\cal A},B) \doteq  (i \slash\!\!\!\partial -
e \,\slash\!\!\!\!\!{\cal A}\, -\, g\, \slash\!\!\!\!B )$.  The 
local gauge invariance of the model emerges from gauge transformations 
acting on the local field subalgebra $\Im$,

$$
\psi^\prime (x) = e^{ \,i\, \,\Lambda (x) }\,\psi (x)\,,
$$
\be
{\cal A}^\prime_\mu (x) = {\cal A}_\mu (x) + \frac{1}{e}\,\partial_\mu \Lambda (x)\,.
\ee

The next step in the functional integral bosonization is to perform 
the decoupling of the  Fermi and vector fields in the Lagrangian (\ref{l1}). To
this end, we introduce the 
parametrization of the vector 
field components $( {\cal A}_{\pm} , B_{\pm} )$ in terms of the 
$U(1)$ group-valued Bose fields $(U_a,V_a)$ and $(U_b,V_b)$ as \cite{BRR},

\be\label{cva}
{\cal A}_+  =  -  \frac{1}{e}\,U_a^{-1}\, i\,\partial _+\,U_a\,\,\,,\,\,\,
{\cal A}_-  =  -  \frac{1}{e}\,V_a\,i\,\partial _-\, V_a^{-1}\,,
\ee

\be\label{cvb}
B_+  =  -  \frac{1}{g}\,U_b^{-1}\, i\,\partial _+\,U_b\,\,\,,\,\,\,
B_-  =  -  \frac{1}{g}\,V_b\,i\,\partial _-\, V_b^{-1}\,.
\ee

\noindent The decoupling is performed by the (Abelian) fermion chiral 
rotation  \cite{FIB,BRR},

\be\label{cfr}
\psi =  {\cal M} \, \chi\,,
\ee

\noindent where the chiral rotation matrix  ${\cal M}$ is given by

\be
{\cal M} = \frac{1}{2}\,(1 + \gamma^5) [U_aU_b]^{-1} +
\frac{1}{2}\,(1 - \gamma^5) V_aV_b\,,
\ee

\noindent in such  way that

\be
\bar \psi\,\slash\!\!\!\!D ({\cal A},B) \psi \, =
\,\chi_{_{\!(1)}}^\dagger\,i\,\partial_-\,\chi_{_{\!(1)}}\,
+ \, \chi_{_{\!(2)}}^\dagger\,i\,\partial_+\, \chi_{_{\!(2)}}\,.
\ee

Introducing in the functional integral the identities,

\bear
1 & = & \int {\cal D} (U_aU_b)\, \det\,(i \partial_+ - e {\cal A}_+ -
g B_+ )\,\delta
\Big (e {\cal A}_+ +  g B_+ - (U_aU_b)^{-1}\,
i\,\partial_+\,(U_aU_b) \Big ) \,,\nonumber\\
1 & = & \int {\cal D} (V_aV_b)\, \det\,( i \partial_- - e {\cal A}_-
- g B_-)\,\delta
\Big ( e {\cal A}_- + g B_- - (V_aV_b)\, i\,\partial_-\,(V_aV_b)^{-1} 
\Big )\,,
\ear

\noindent the change of variables from $({\cal A}_+,{\cal A}_-)$ 
to $(U_a,V_a)$ and $(B_+,B_-)$ 
to $(U_b,V_b)$, is performed integrating over the vector 
fields components  $( {\cal A}_\pm , B_\pm )$. Performing the 
fermion chiral rotation (\ref{cfr}) and
taking into account the  corresponding change in the 
integration measure \cite{FIB,BRR}, we get

\be
{\cal D} \bar \psi {\cal D} \psi\,
{\cal D} {\cal A}_\pm\, B_\pm  =
\,{\cal D} \bar \chi {\cal D} \chi\,
{\cal D} U_a\,{\cal D} U_b \,{\cal D} V_a\,{\cal D} V_b\,
\mbox{\Large${\cal J}$}[U_a,U_b,V_a,V_b]\,,
\ee

\noindent with 

\be\label{jacobian}
\mbox{\Large${\cal J}$} \,=\,e^{\textstyle\,- i \,\Big ( \Gamma [U_aU_b] + 
\Gamma [V_aV_b]\,+\, c\,\int\,d^2 z\,
 [(U_a U_b)^{-1} \partial_+ (U_aU_b)] \,
[(V_a V_b) \partial_- (V_a V_b)^{-1}]\,\Big )}\,,
\ee

\noindent where $\Gamma [G]$ is the  Wess-Zumino-Witten (WZW) functional
\cite{WZW}, which enters in (\ref{jacobian}) with negative level. In the 
abelian case the WZW functional reduces to the free action

\be
\Gamma [G] =  \Gamma [G^{-1}] = \frac{1}{8 \pi}\,\int\,d^2 z\, 
\partial _\mu G \partial ^\mu G^{-1}\,.
\ee

Using the Abelian Polyakov-Wiegman identity \cite{PW},

\be\label{PW}
\Gamma [U V] = \Gamma [U] + \Gamma [V] + \frac{1}{4 \pi}\,\int\,d^2 z\, \Big (
U^{-1} \partial_+ U \Big ) \Big ( V \partial_- V^{-1} \Big )\,,
\ee

\noindent and the gauge invariant regularization $ c = 1 / 4 \pi$, we can write
(\ref{jacobian}) in terms of the gauge invariant 
Bose fields $ G_\alpha \doteq U_\alpha V_\alpha $ ( $\alpha = a,b. )$, 

\be
{\cal J} = e^{\textstyle\,- i\,\Gamma [G_a G_b]}\,.
\ee

\noindent The Maxwell Lagrangian can be written as

\be
- \frac{1}{4} {\cal F}_{\mu \nu}^2 = \frac{1}{8 e^2} \Big \{
\partial_+ \Big ( G_a i \partial_- G_a^{-1} \Big ) \Big \}^2\,,
\ee

\noindent and the  total effective action is given by

$$
S _{_{eff.}}^\prime =\, - \,\Gamma [ G_a G_b]\,+  
 \,\int\,d^2 z\, \,\Bigg \{\,\frac{1}{8 e^2} \Big [
\partial_+ \Big ( G_a i \partial_- G_a^{-1} \Big ) \Big ]^2\,
 -\,\frac{1}{2 g^2}\,
\,\Big ( U_b^{-1} \partial_+ U_b \Big )
\Big ( V_b \partial_- V_b^{-1} \Big )\,
+
$$ 

\be\label{Seff}
+\,\bar \chi\, i\,\slash\!\!\!\partial\,\chi\,-\,
M_o\, \Big ( \chi^\ast_{_{\!(1)}} \chi_{_{\,(2)}}\,(G_a G_b)^{-1} +
\chi^{\ast}_{_{\!(2)}} \chi_{_{\,(1)}}\,( G_a  G_b)\,\Big ) \,
\Bigg \}\,
\ee

The vector fields in two-dimensions can be decomposed as

\be\label{A}
{\cal A}_\mu =  \frac{\sqrt \pi}{e}\,( \epsilon_{\mu \nu}\,\partial^\nu\,
 \phi_a + 
\partial _\mu \xi_a )\,,
\ee

\be\label{B}
B_\mu =  \frac{\sqrt \pi}{g}\,( \epsilon_{\mu \nu}\,\partial^\nu\,
\phi_b + \partial _\mu \xi_b )\,,
\ee

\noindent which corresponds to parametrizing the Bose fields $(U_\alpha,V_\alpha)$
as follows,

\be
U_\alpha = e^{\textstyle\, i\,\sqrt \pi\, (\phi_\alpha + \xi_\alpha)}\,,
\ee

\be
V_\alpha = e^{\textstyle\,i\,\sqrt \pi\, ( \phi_\alpha - \xi_\alpha)}\,.
\ee

\noindent The effective Lagrangian density, corresponding to the action
(\ref{Seff}), can be written as

$$
{\cal L}_{_{eff.}}^\prime = 
\frac{\pi}{2 g^2}\,(\partial_\mu \xi_b )^2\,
 +\,\frac{1}{2m^2}\,(\Box  \phi_a )^2\,+\,
\frac{1}{2}  \phi_a \Box \phi_a + \frac{1}{2} ( 1 +
\frac{\pi}{g^2} )  \phi_b
\Box  \phi_b +   \phi_a \Box \phi_b\,+
$$

\be\label{leff}
+ \bar \chi\, i\,\slash\!\!\!\partial\,\chi\,-\,
M_o\, ( \chi^\ast_{_{\!(1)}} \chi_{_{\,(2)}}\,e^{\,\textstyle
- 2 i\,\sqrt \pi\,( \phi_a +  \phi_b) } + \mbox{h. c.}\,)\,,
\ee

\noindent where $m^2 = e^2 /\pi $, is the mass of the gauge field in the 
absence of the 
Thirring coupling ($g = 0$, standard $QED_2$). Due to the 
local gauge invariance of the
model, the field $\xi_a$ does not appears in the effective 
Lagrangian (\ref{leff}). The field $\xi_a$ is an unphysical 
pure gauge excitation, that appears only in the gauge non-invariant field 
operators $\psi$ and ${\cal A}_\mu$, and can be gauged away from the operator 
solution (\ref{cva}-\ref{cfr}). However, as we shall see, in the anomalous 
chiral model the field $\xi_a$ acquires a non trivial dynamics and plays an 
important role in the algebraic structure of the model.

Defining the canonical field 

\be
\phi_b^\prime = \frac{\sqrt \pi}{g} (1 + \frac{g^2}{\pi}
)^{1/2} \phi_b = \frac{\sqrt \pi}{g} \alpha^{-1/2}  \phi_b\,,
\ee

\noindent the fields ($\phi_a,\phi_b$) can be decoupled using the identity

$$
 \phi_b^\prime\, \Box\,  \phi_b^\prime \,+\, 2\,
 \frac{g}{\sqrt \pi}\, \alpha^{1/2}\, (\Box\, \phi_a )\,
   \phi_b^\prime\, =
$$

$$
=( \phi_b^\prime +  \frac{g}{\sqrt \pi}\,
 \alpha^{1/2}\,  \phi_a )\, \Box\,
( \phi_b^\prime +  \frac{g}{\sqrt \pi}\, \alpha^{1/2}\,
 \phi_a )\, -\, \frac{g^2}{\pi}\, \alpha\,
  \phi_a \,\Box\, \phi_a \,=
$$

\be
=\,  \eta_b \,\Box\,  \eta_b\,
 -\, \frac{g^2}{\pi}\, \alpha \, \phi_a \Box \phi_a\,,
\ee

\noindent where we defined the new field

\be
\eta_b =  \phi_b^\prime + \frac{g}{\sqrt \pi}
\,\alpha^{1/2}\,  \phi_a\,.
\ee

\noindent Introducing the parameter

\be
\beta^2 = \frac{4 \pi}{(1 + \frac{g^2}{\pi})} = 4 \pi \alpha\,,
\ee

\noindent and rescalling the fields 

\be
\phi_a^\prime = \frac{\beta}{2 \sqrt \pi}\,\phi_a\,\,\,\,,\,\,\,\,
\xi_b^\prime = \frac{\sqrt \pi}{g} \,\xi_b\,,
\ee

\noindent the effective Lagrangian density can be written as

$$
{\cal L}_{_{eff.}}^\prime = 
\frac{1}{2 }\,(\partial_\mu \xi^\prime_b )^2\,
- \frac{1}{2} (\partial_\mu \eta_b )^2\,
 +\,\frac{1}{2m_\ast^2}\,(\Box \phi_a^\prime )^2\,+\,
\frac{1}{2}  \phi_a^\prime \Box  \phi_a^\prime + 
$$

\be\label{lefff}
+ \bar \chi\, i\,\slash\!\!\!\partial\,\chi\,-\,
M_o\, ( \chi^\ast_{_{\!(1)}} \chi_{_{\,(2)}}\,e^{\,\textstyle
- \, i\,\beta\,(\frac{g}{\sqrt \pi}\, \eta_b + \phi_a^\prime) } + 
\mbox{h. c.}\,)\,,
\ee

\noindent where $m_\ast^2$ is the mass of the gauge field ${\cal A}_\mu$ in the
presence of the Thirring coupling \cite{Dutra},

\be
m_\ast^2\, =\, m^2\, \frac{\beta^2}{4 \pi} = \frac{e^2}{\pi + g^2}\,.
\ee

In order to decompose the higher derivative term in
(\ref{lefff}) in terms of a general local solution, we enlarge the Bose field algebra
by using the functional integral identity 

$$
\int {\cal D} \phi_a^\prime\,e^{\textstyle i \int d^2 z \Big \{
\frac{1}{2 m_\ast^2} \,\phi_a^\prime\,\Box^2\,\phi_a^\prime + \frac{1}{2}\,
\phi_a^\prime \Box \phi_a^\prime \Big \}} \equiv
$$

$$
\equiv \int {\cal D} A\, \int {\cal D} \phi_a^\prime\,e^{\textstyle i 
\int d^2 z \Big \{ - \frac{1}{2}\,A^2 + 
\frac{1}{m_\ast} (\Box A) \phi_a^\prime  + 
\frac{1}{2}  \phi_a^\prime
\Box  \phi_a^\prime \Big \}} \,=
$$

\be
=\, \int {\cal D}  \Sigma \int {\cal D} \eta_a\,e^{\textstyle i 
\int d^2 z \Big \{ - \frac{1}{2}\, \Sigma \Big ( \Box + m_\ast^2 \Big )
 \Sigma - \frac{1}{2} (\partial_\mu \eta_a )^2 \Big \}}
\,,
\ee

\noindent where we have defined the fields

\be
\eta_a =  \phi_a^\prime + \frac{1}{m_\ast} A\,\,\,\,,\,\,\,\,
 \Sigma = \frac{1}{m_\ast} A\,.
\ee 
 
\noindent The effective Lagrangian (\ref{lefff}) can be written as

$$
{\cal L}_{_{eff.}}^\prime = 
\frac{1}{2}\,(\partial_\mu \xi^\prime_b )^2\,
- \frac{1}{2} (\partial_\mu \eta_a )^2\,
- \frac{1}{2} (\partial_\mu \eta_b )^2\,
+ \frac{1}{2} (\partial_\mu  \Sigma )^2\,
- \frac{1}{2} m_\ast^2  \Sigma^2\,+ 
$$

\be\label{leffff}
+ \bar \chi\, i\,\slash\!\!\!\partial\,\chi\,-\,
M_o\, ( \chi^\ast_{_{\!(1)}} \chi_{_{\,(2)}}\,e^{\,\textstyle
-  i\,\,\beta ( \eta_a + \frac{g}{\sqrt \pi}\,\eta_b 
 -   \Sigma) } + h. c.\,)\,.
\ee

The last step is to perform the bosonization of the Fermi fields $\chi (x)$. In
terms of the Bose fields, the original Fermi field operator $\psi (x)$ can 
be written as

\be
\psi (x) \,=\, \,:e^{\, -\, i\, \frac{\beta}{2}\,
\gamma^5\,  \Sigma (x)}:\,
:e^{\, i\, \frac{\beta}{2}\, \gamma^5\,
[\,\eta_a (x)\, + \,\frac{g}{\sqrt \pi}\,\eta_b (x)\,]}:\,
:e^{ \,-\, i\, g\, \xi^\prime_b (x)}:\, \chi (x)\,.
\ee

The $2n$-point correlation functions of 
the Fermi fields are obtained by functional derivation of the 
generating functional with respect to the Grassmann valued 
sources $\bar \upsilon$ and $\upsilon$, and can be written as

$$
\langle \Omega \vert\,\bar \psi (x_1)\,\cdots\,\bar \psi (x_n)\,
\psi (y_1)\,\cdots\,\psi (y_n) \vert \Omega \rangle = 
\langle \Omega \vert \prod_{i = 1}^n\,e^{\, i\, g \,\xi_b^\prime (x_i)}\,
\prod_{j = 1}^n\,e^{\, -\, i\, g\, \xi_b^\prime (y_j)}\,\vert \Omega 
\rangle_o\,\times
$$

\be\label{cf}
 \langle \Omega \vert \prod_{i = 1}^n\,\bar \chi (x_i)\,e^{ i \frac{\beta}{2} 
\gamma^5 [\, \eta_a (x_i) + \frac{g}{\sqrt \pi}\,\eta_b
 (x_i) -  \Sigma (x_i)]}\,
\prod_{j =1}^n\,
\chi (y_j)\,e^{ - i \frac{\beta}{2} 
\gamma^5 [\, \eta_a (y_j) + \frac{g}{\sqrt \pi}\,  
\eta_b  (y_j) -  \Sigma (y_j)]}\,
\vert \Omega \rangle_I\,,
\ee

\noindent where the notation $\langle \Omega \vert \cdots \vert \Omega \rangle_o$ means average with 
respect to the free massless $\xi_b^\prime$-field theory, with the functional
integral measure

\be
d \mu (\xi^\prime_b) = {\cal D}\,\xi^\prime_b\,e^{\,i\,
S_o [\xi^\prime_b]}\,,
\ee

\noindent and $\langle \Omega \vert \cdots \vert \Omega \rangle_I$ means average
with respect to the  effective 
Lagrangian  
${\cal L}_I [\bar \chi,\chi, \eta_a, \eta_b,  \Sigma]$, with the measure,

\be
d \mu_{_{\!I}} = {\cal D}\,\bar \chi\,{\cal D}\,\chi\,{\cal D}\,\eta_a\,
{\cal D}\,\eta_b\,{\cal D}\,\Sigma\,e^{\,i\,\int\,d^2 z\,{\cal L}_I [\bar \chi,\chi,\eta_a, \eta_b,  \Sigma]
}\,.
\ee

\noindent   Following the standard 
procedure \cite{FIB,Naon,Fro}, we perform the expansion
of the exponential of the interaction term 
of  ${\cal L}_I $ in a
power series of the bare mass $M_o$, which provides the functional integral
Gell-Mann and Low formula,

$$
d \mu_{_{\!I}} = \sum_{n=0}^\infty\,(- i M_o)^n \frac{1}{n!}\,\prod_{j=1}^n\,
\int d^2 z_j\, {\cal D}\,\bar \chi\,{\cal D}\,\chi\,{\cal D}\,\eta_a\,
{\cal D}\,\eta_b\,{\cal D}\,\Sigma\,
e^{\,i\,S_o [\Sigma]}\,
e^{\,-\,i\,S_o [\eta_a]}\,
e^{\,-\,i\,S_o [\eta_b]}\,
e^{\,i\,S_o [\bar \chi,\chi]}\,\times
$$

\be
\prod_{k=1}^n\,\Big [\,
 \chi^\ast_{_{\!(1)}}(z_k) \chi_{_{\,(2)}} (z_k)\,e^{\,\textstyle
-  i\,\,\beta \{ \eta_a(z_k) + \frac{g}{\sqrt \pi}\,\eta_b(z_k)
 -   \Sigma(z_k)\} } + \mbox{h. c.} \Big ]\,.
\ee
 The resulting correlation 
function, besides
the contributions of the 
fields $ \eta_a$, $ \eta_b$ and $ \Sigma$, corresponds
to averages of products of chiral density 
operators $\bar \chi (x_i) \chi (y_j)$ with respect to
the free massless Fermi theory. The effective bosonized theory can be obtained by
reconstructing the series using the free field bosonization
expressions 

\be\label{ff}
\chi (x) = \Big ( \frac{\mu_o}{2 \pi} \Big )^{1/2}\,
e^{\textstyle \, - i \frac{\pi}{4}\,\gamma^5}\,
:e^{{\textstyle
i\,\sqrt \pi \{\,\gamma^5\, \varphi (x) + }\int_{x^1}^{+ \infty}\,
\textstyle \dot{\varphi} (x^0,z^1)\,d z^1 \}}:\,,
\ee

\be
\bar \chi\, i\,\slash\!\!\!\partial\,\chi \equiv 
\frac{1}{2}\,: (\partial_\mu \varphi )^2 :\,,
\ee

\be
\chi_{_{\!(1)}}^\ast \chi_{_{\!(2)}} \equiv \frac{\mu_o}{2 \pi}\,
: e^{\,\textstyle 2 i \sqrt \pi \varphi }:\,,
\ee

\noindent where the double dots indicate normal ordering with respect to the 
free propagator $(\Box + \mu_o^2)^{\,-1}$ in the limit $\mu_o \rightarrow 0$. In 
this way, the effective bosonized Lagrangian density is given by

$$
{\cal L}_{_{eff.}}^\prime = 
\frac{1}{2}\,(\partial_\mu \xi^\prime_b )^2\,
- \frac{1}{2} (\partial_\mu \eta_a )^2\,
- \frac{1}{2} (\partial_\mu \eta_b )^2\,
+ \frac{1}{2} (\partial_\mu  \Sigma )^2\,
- \frac{1}{2} m_\ast^2  \Sigma^2\,+ 
$$

\be\label{lag}
+ \frac{1}{2} (\partial_\mu \varphi )^2 + M \cos \{
2 \sqrt \pi  \varphi - \beta (\eta_a + \frac{g}{\sqrt \pi} \eta_b -
 \Sigma )\}\,.
\ee

We introduce two independent fields $ \Phi$ and $ \zeta$, through the 
canonical transformation,

\bear
\gamma  \Phi & = &  2 \sqrt \pi \varphi -
\frac{g}{\sqrt \pi}\, \beta \,\eta_b\,,\\
\gamma \zeta  & = &  \frac{g}{\sqrt \pi} \beta \varphi
- 2 \sqrt \pi \,\eta_b\,,
\label{ct}
\ear

\noindent with 

\be
\gamma^2 = 4 \pi - \frac{g^2}{\pi} \beta ^2  = \beta^2\,
\ee

\noindent  and the field $\zeta$ is quantized with negative metric. In 
terms of these new fields, the effective bosonized
Lagrangian density can be written as

$$
{\cal L}_{_{eff.}} = 
\frac{1}{2}\,(\partial_\mu \xi^\prime_b )^2\,
-\, \frac{1}{2} (\partial_\mu \zeta )^2
\,-\, \frac{1}{2} (\partial_\mu \eta_a )^2\,
+ \,\frac{1}{2} (\partial_\mu  \Phi )^2\,+
$$

\be\label{sg}
+ \,\frac{1}{2} (\partial_\mu  \Sigma )^2\,
- \,\frac{1}{2} m_\ast^2  \Sigma^2\,
 +\, M \cos \{
\beta  \Sigma + \beta (  \Phi \,-\,\eta_a) \}
\,.
\ee

\noindent The effective bosonized theory is described by three coupled sine-Gordon-like
fields $ \Sigma$, $ \Phi$ and $ \eta_a$, with $ \eta_a$ quantized
with negative metric, and two ``decoupled'' free massless 
fields $\xi^\prime_b$ and $\zeta$, quantized with opposite metric. The
field $ \Sigma$ is a massive sine-Gordon field with mass
$m^2_\ast = m^2 \beta^2 / 4 \pi$. The local gauge invariance of the
model ensures that the Lagrangian (\ref{sg}) carries no dependence
on the Bose field $\xi_a$, which is a pure gauge excitation. However, as we shall 
see, in the anomalous model the
field $\xi_a$ plays an important role in the field 
algebra. 

%%%%%%%%%%%%%%%%%%%%%%%%%%%%%%%%%%%%%%%%%
\subsection{Field algebra and Hilbert space}
%%%%%%%%%%%%%%%%%%%%%%%%%%%%%%%%%%%%%%%%%%%%%

The introduction of the auxiliary vector field $b_\mu$ in the functional 
integral
bosonization leads to an 
enlarged Bose field algebra containing two ``decoupled'' massless 
fields $\xi^\prime_b$ and $\zeta$, quantized with opposite metric. As a 
matter of fact,  the extraction of 
these decoupled massless Bose fields relies on a structural  problem 
which is related to the fact that the fields $\xi^\prime_b$  and $\zeta$ do not
belong to the field algebra $\Im^\prime$ and  cannot be defined by
itself as operators in the Hilbert space ${\cal H}^\prime$ \cite{FIB,M,Bel,BRR}. We 
shall return to this point further on.

In order to display the structure of the Hilbert spaces ${\cal H}^\prime$ 
and ${\cal H}$, it is instructive to express the  
fields $\{\bar \psi, \psi, {\cal A}_\mu, B_\mu \}$, and the 
corresponding source terms in the generating functional (\ref{mgf}), in terms
of the set of 
Bose fields $\{ \Sigma,  \Phi, \eta_a,  \xi^\prime_b,  \zeta \}$. Using the 
decomposition (\ref{B}) for the auxiliary vector field $B_\mu$, and the corresponding Bose field
transformations, we get

\be B_\mu\, (= \,g J_\mu \,+\, b_\mu)\,
= \, - \,g\,\frac{\beta}{2 \pi}\,\epsilon_{\mu \nu} \partial ^\nu\,
 \Big \{\,  \Sigma\,+\, \Phi\,-\,\eta_a\,\Big \}\,
 +\, \partial_\mu\,(\,\xi_b^\prime - \zeta \,)\,.
\ee

\noindent In this way, the vector current $J^\mu$ is identified as being given
by

\be
J_\mu \,=\, - \,\frac{\beta}{2 \pi}\,\epsilon_{\mu \nu}\, \partial ^\nu\,
  \Sigma\,+ L_\mu\,,
\ee

\noindent where

\be
L_\mu \,=\, -\, \frac{\beta}{2 \pi}\,\epsilon_{\mu \nu} \partial ^\nu (  \Phi\,
 -\,\eta_a )\,\equiv\,\epsilon_{\mu \nu} \partial ^\nu  L\,,
\ee

\noindent is a longitudinal current. Since  the 
sine-Gordon fields $ \Phi$ and $ \eta_a$ are quantized with 
opposite metric, the corresponding equations of motion are given by 

\be
\Box \Phi - \beta\,M\,:\sin\,\{\beta \Sigma + \beta (\Phi\,-
\,\eta_a)\}: = 0\,,
\ee

\be
-\,\Box \eta_a + \beta\,M\,:\sin\,\{\beta \Sigma + \beta (\Phi\,-
\,\eta_a)\}: = 0\,.
\ee

\noindent In this way, the field 
combination $ L = ( \Phi - \eta_a)$, is a free
massless field

\be
\Box (  \Phi -  \eta_a ) = 0\,,
\ee

\noindent which generates zero norm states from the vacuum,

\be
\langle \Omega \vert\,\Big (  \Phi (x) -  \eta_a (x) \Big ) \Big (
 \Phi (y) -  \eta_a (y) \Big )\,\vert \Omega \rangle = 0\,.
\ee

The auxiliary vector field $b_\mu$ is identified with a
longitudinal current $\ell_\mu$, given in terms of the two free and massless Bose fields
quantized with opposite metric

\be\label{lc}
b_\mu \equiv \ell_\mu = \, \partial _\mu ( \xi^\prime_b - \zeta )\,,
\ee

\noindent such that

\be
\langle \Omega \vert \mbox{\large{$\ell$}} _\mu (x)\,
\mbox{\large{$\ell$}} _\mu (y)\, \vert \Omega \rangle^\prime  = 0\,.
\ee

\noindent In the limit $g = 0$, the vector field $B_\mu$ is a longitudinal
field $ B_\mu \equiv \ell_\mu $.

In a similar way, the gauge field (\ref{A}) can be written as

\be
{\cal A}_\mu = -\,\frac{1}{m_\ast}\,\epsilon_{\mu \nu} \partial ^\nu ( \Sigma
- \eta_a )\,.
\ee

Performing the 
fermion chiral rotation (\ref{cfr}), together with the canonical 
transformations (\ref{ct}), and using the
bosonized expression (\ref{ff}) for the free massive Fermi field, we obtain
the operator solution for the Fermi field $\psi (x)$ of the massive 
Schwinger-Thirring model in terms of 
the Mandelstam \cite{Mandelstam} ``soliton'' 
field operator as 

\be\label{psi}
\psi (x) = \Psi (x) \,\Upsilon (x)\,,
\ee

\noindent where $\Psi$ is the operator solution of the
massive Schwinger-Thirring  model 

\be
\Psi (x)\, =\, :e^{\, i \,\frac{\beta}{2}\, \gamma^5\,\{
 \Sigma (x)\,-\,\eta_a (x)\, \}}:\,{\cal S} (x) \,,
\ee

\noindent where ${\cal S} (x)$ is the Mandelstam representation for the massive
Fermi field operator of the Thirring model \cite{Mandelstam}

\be
{\cal S} (x) = \Big ( \frac{\mu_o}{2 \pi} \Big )^{1/2}\, 
e^{\,-\, i\, \frac{\pi}{4} \gamma^5} :e^{\, i\, \frac{\beta}{2} \gamma^5 \,
 \Phi (x)\,+\,\frac{2 \pi}{\beta}\,i\,\int_{x^1}^\infty\,
{\dot{ \Phi}} (x^0,z^1) dz^1 }:\,.
\ee

\noindent  In Eq. (\ref{psi}), both spinor 
components of the Fermi field  $\Psi_{_{\!(\alpha)}}(x)$  are 
multiplied by the exponential field 

\be\label{spurious}
\Upsilon (x)\, =\, :e^{\,-\,i\, g\, \{\,\xi_b^\prime (x)\,-\,\zeta (x)\, \}}:\,=\,
 : e^{\,i g \,\mbox{\large{$\ell$}} (z) } :\,,
\ee

\noindent where $\mbox{\large{$\ell$}} = \xi^\prime_b - \zeta$, is the
potential for the longitudinal current (\ref{lc}).

Besides the electric field and the vector current, the gauge invariant field 
subalgebra is generated by the bilocal operators formally defined 
by \cite{LS,RS,S,B,AAR}

\be\label{bilocal}
D_{\alpha \beta} (x,y) \approx \psi_\alpha^\ast (x)\,e^{\,i e \int_y^x {\cal A}_\mu (z) d z^\mu}\,
\psi_\beta (y)\,.
\ee

\noindent The gauge invariant composite operator (\ref{bilocal}) can be 
factorized as

\be\label{fbilocal}
D_{\alpha \beta} (x,y)\,=\,T_{\alpha \beta} (x,y)\,
\sigma^\ast_\alpha (x) \sigma_\beta (y)\,W_{_{\!b}} (x,y)\,,
\ee

\noindent where $T_{\alpha \beta} (x,y)$ is the generalization of
the bilocal operator obtained in \cite{RS} for the massive $QED_2$, 

\be
T_{\alpha \beta} (x,y)\,=\,
N (x - y)\,:e^{\,-\,i\,\frac{\beta}{2}\,\{\, \gamma^5_{\alpha\alpha}\, 
\Sigma (x)\,-\,\gamma^5_{\beta\beta}\, \Sigma (y)\, \}\,-\,
\frac{2 \pi}{\beta}\,i\,\int_y^x\,\epsilon_{\mu \nu} \partial^\nu 
 \Sigma (z) d z^\mu} :\,,
\ee

\noindent $\sigma_\alpha (x)$ is the spurious operator introduced by
Lowenstein and Swieca \cite{LS,RS,S},

\be\label{lsso}
\sigma_\alpha (x)\,=\,
:e^{\,i\,\textstyle \frac{\beta}{2}\,\gamma^5_{\alpha\alpha}\,\{\, \Phi (x)\,-\,
\eta_a (x)\,\}\,-\,\frac{2 \pi}{\beta}\,i\,\int_{x^1}^\infty\,\{\,
\dot{ \Phi} (z)\,-\,\dot{\eta}_a (z) \}\,d z^1 }:\,,
\ee

\noindent and $W_{_{\!b}} [x,y]$ is a spurious ``bilocal''operator given by

\be\label{w}
W_{_{\!b}} [x,y]\,\doteq\,
:e^{\textstyle\,i g \int_{x,C}^{y} \ell_\mu (z) d z^\mu}:\,,
\ee

\noindent where $C$ is an arbitrary curve.

The chiral density operator is computed from the point-splitting limit 
procedure of the bilocal operator (\ref{bilocal}), and we get

\be\label{J}
{\cal J} (x) \doteq \vdots \psi_{_{\!(1)}}^\ast (x)\,\psi_{_{\!(2)}}
(x) \vdots =
: e^{\,- i \beta\, \Sigma (x)}:\,\sigma_{_{\!(1)}}^\ast (x)
\sigma_{_{\!(2)}} (x)\,,
\ee

\noindent where

\be
\sigma_{_{\!(1)}}^\ast (x)\,\sigma_{_{\!(2)}} (x) = :e^{\,i \beta [
 \Phi (x) - \eta_{_{a}} (x)]}:\,,
\ee

\noindent is a spurious operator carrying the free fermion 
chirality \cite{LS,RS,S,B}.

The interpolating generating functional (\ref{mgf}), from which we construct the 
Hilbert space ${\cal H}^\prime$, can be written in terms of the Bose fields as

$$
{\Z}^{\prime} [\upsilon,\bar \upsilon,\omega_\mu,\rho_\mu] = 
\langle e^{\,\textstyle i\,\int d^2z\,(\bar \psi \upsilon + \bar \upsilon \psi + {\cal A}_\mu
\omega^\mu + b_\mu \rho^\mu )} \rangle^\prime =
$$

\be\label{fbgf1}
= \langle e^{\,\textstyle i\,\int d^2z\,( \overline \Psi \Upsilon^\ast \upsilon + \bar \upsilon
\Psi \Upsilon + 
\frac{1}{m_\ast}\,\omega^\mu \epsilon_{\mu \nu} \partial ^\nu ( \Sigma
- \eta_a )\,+\,\rho^\mu \partial_\mu [\xi^\prime_b - \zeta] )} 
\rangle^\prime
\ee

\noindent where the average is taken with respect to the measure

\be
d \mu^\prime = {\N}^{-1}\,
\int\,{\cal D} \xi_b^\prime\,
e^{\,\textstyle \,i\, S_o [\xi_b^\prime]}\,
\int\,{\cal D} \zeta\,
e^{\,\textstyle \,- i\, S_o [\zeta]}\,
\int\,{\cal D}  \Phi\,\int\, \,{\cal D} \eta_a\,\int\,
{\cal D}  \Sigma\,\,
e^{\,\textstyle i\, S\, [ \Sigma,
\Phi, \eta_a]}\,, 
\ee

\noindent where the $S_o$'s are the free actions for the
massless fields quantized with opposite metric.

>From the  generating functional (\ref{fbgf1}) we obtain the 
general Wightman $2n$-point functions for the Fermi field in terms of 
averages of order-disorder operators 

$$
\langle \Omega \vert\,\bar \psi (x_1)\,\cdots\,\bar \psi (x_n)\,
\psi (y_1)\,\cdots\,\psi (y_n) \vert \Omega \rangle^\prime =  
$$
\be\label{ffwf}
\langle \Omega \vert\,\overline{\Psi} (x_1)\,\cdots\,
\overline{\Psi} (x_n)\,\Psi (y_1)\,
\cdots\, \Psi (y_n)\,\vert \Omega \rangle \langle \Omega \vert \Upsilon^\ast (x_1)\,
\cdots\,\Upsilon^\ast (x_n)\,\Upsilon (y_1)\,\cdots\, \Upsilon (y_n)\, 
\vert \Omega \rangle_o\,,
\ee

\noindent where the notation $\langle \Omega \vert \cdots \vert \Omega
\rangle$ means average with respect to coupled sine-Gordon 
fields $\Sigma$, $\Phi$, $\eta_a$, and $\langle \Omega \vert \cdots \vert \Omega \rangle_o$
 means average with respect to the free theories of the massless 
Bose fields $\xi^\prime_b$ and $\zeta$.  Due to the opposite metric 
quantization for the fields $\zeta$ and $\xi^\prime_b$, the functional 
integration over the field $\xi^\prime_b$ cancels those arising from the 
integration over the field $\zeta$. This implies that the field $W_{_{\!b}}$
generates constant constributions to the Wightman functions,

$$
\langle \Omega \vert \, \Upsilon^\ast (x_1)\,\cdots\,\Upsilon^\ast (x_n)\,
\Upsilon (y_1)\,\cdots\,\Upsilon (y_n)\, \vert \Omega \rangle_o \equiv
\langle \Omega \vert \prod_{j = 1}^{n} :e^{\textstyle\,i g 
\int_{x_j,C_j}^{y_j} \ell_\mu (z) d z^\mu}: \vert \Omega \rangle_o 
= 
$$

\be\label{cd}
=\,\langle \Omega \vert \prod_{j = 1}^n\,W_{_{\!b}} [x_j,y_j] \vert 
\Omega \rangle_o = 1\,.
\ee

\noindent  The fact that the operator $W_{_{\!b}} [x,y]$ generates 
constant constributions in (\ref{ffwf}), implies the 
isomorphism between the 
Wightman functions of the Fermi field in the Hilbert 
spaces ${\cal H}^\prime$ and ${\cal H}$:
\be\label{iso}
\langle \Omega \vert\,\bar \psi (x_1)\,\cdots\,\bar \psi (x_n)\,
\psi (y_1)\,\cdots\,\psi (y_n) \vert \Omega \rangle^\prime \equiv  
\langle \Omega \vert\,\overline{\Psi} (x_1)\,\cdots\,
\overline{\Psi} (x_n)\,\Psi (y_1)\,
\cdots\, \Psi (y_n)\, \vert \Omega \rangle\,.
\ee

\noindent For any functional ${\cal F} \{\bar \psi,\psi\} \in \Im^\prime$,
we obtain the general one-to-one mapping in the Hilbert
spaces ${\cal H}^\prime$ and ${\cal H}$:
\be\label{iso1}
\langle \Omega \vert\,{\cal F}\{\bar \psi, \psi\}\, \vert \Omega \rangle^\prime \equiv
\langle \Omega \vert\,{\cal F}\{\overline{\Psi},\Psi\}\,
\vert \Omega \rangle\,.
\ee

The enlarged Bose field algebra $\Im^{^{B}}$ contains two spurious 
fields with zero scale dimension, e. g., $\Upsilon (x)$ and 

\be
\hat \sigma (x) \doteq \sigma_{_{\!(1)}}^\ast (x)\,\sigma_{_{\!(2)}} (x)\,.
\ee

\noindent As is well known from the $QED_2$ with massless 
Fermi fields \cite{RS,B}, the
spurious operator $\hat \sigma (x)$ is the generator of the integer winding number gauge
transformations in the physical Hilbert space. This will remains 
valid in the present model. However, the operator $\Upsilon (x)$ cannot be 
defined by 
itself in the field algebra $\Im^\prime$ \cite{FIB}. Nevertheless, the operator $W_{_{\!b}} (x,y)$ belongs
to the field algebra $\Im^\prime$, but it reduces to the identity
in ${\cal H}^\prime$. From the 
functional integral point of view, although the 
partition function obtained from (\ref{fbgf1}) 
factorizes in the form 

\be
{\Z}^\prime [0] = {\Z}^o_{_{\zeta}} [0]\times
{\Z}^o_{_{\!\xi^\prime_b}} [0] \times {\Z}_{_{\!\Sigma,\Phi,\eta_a}}[0]\,,
\ee

\noindent the fact that the 
`` spurious ''  field $\Upsilon$ appears  attached to the bosonized Fermi 
field $\Psi$ in the 
source terms, suggests that the  generating functional (\ref{fbgf1}) cannot be
factorized and the so-called ``decoupled'' massless scalar fields cannot be 
removed in a naive way, contrary to what is usually 
done \cite{Carena,Naon}. The 
origin of this extra spurious field relies on the 
functional integral bosonization procedure that makes use of the
introduction of the auxiliary vector field $B_\mu$. This question can 
be clarified on the basis of the 
intrinsic algebraic structure of the model.

The set of fields $\{\bar \psi,\psi,{\A}_\mu\}$ constitute the 
intrinsic mathematical 
structure of the model and  generate the local polynomial field 
algebra $\Im = \Im \{\bar \psi,\psi,{\A}_\mu\}$. The
Wightman functions generated from the field algebra $\Im$ define the model and
identifies the Hilbert space ${\cal H}$ of the 
theory, $ {\cal H} \doteq \Im \Omega$. The introduction of
the auxiliary vector field $b_\mu$ enlarge the field algebra $\Im \rightarrow
\Im^\prime = \Im^\prime \{b_\mu,\bar \psi,\psi,{\A}_\mu\}$, and the change 
of variables 
(\ref{cv1}) leads to a new field 
algebra $\Im^\prime = \Im^\prime \{{\B}_\mu,\bar \psi,\psi,{\A}_\mu\}$. This
field algebra is represented in the enlarged Hilbert 
space ${\cal H}^\prime \doteq \Im^\prime \Omega$.  Within the 
bosonization procedure the fundamental fields defining the field algebra
$\Im^\prime$ are written in terms  of the Bose
fields $\{ \Sigma, \Phi, \eta_a,\zeta,\xi^\prime_b \}$. This set of  Bose 
fields define an
enlarged redundant field algebra $\Im^{^{B}}$, which is represented 
in the indefinite metric Hilbert 
space ${\cal H}^{^B} \doteq \Im^{^B} \Omega$. These Bose fields are the building blocks
in terms of which the bosonized  solution is constructed and, as 
stressed in Refs. \cite{M,Bel,BRR}, should not
be considered as elements of the field 
algebra $\Im^\prime$. Only some particular combinations of them belong to 
the field algebra
$\Im^\prime$, in such a way that, $ \Im^\prime \subset \Im^{^{\!B}}$, and 
thus, $ {\cal H}^\prime \subset {\cal H}^{^B}$. The auxiliary 
vector field $B_\mu = g J_\mu + \mbox{\large{$\ell$}}_\mu $, belong 
to the field algebra $\Im^\prime$ and since $ J_\mu \in \Im^\prime$, then, $\mbox{\large{$\ell$}}_\mu \in \Im^\prime$. In this way, the positive 
semi-definite Hilbert space ${\cal H}^\prime$ is generated from the field 
algebra $\Im^\prime \{B_\mu,\bar \psi,\psi,{\A}_\mu \} = 
\Im^\prime \{ \Im^\prime_o,\overline{\Psi} \Upsilon^\ast,\Psi \Upsilon, {\A}_\mu 
\}$, where $\Im^\prime_o \subset \Im^\prime$ is the field subalgebra 
generated by the
longitudinal current $\mbox{\large{$\ell$}}_\mu$, $\Im^\prime_o 
= \Im^\prime_o\{\mbox{\large{$\ell$}}_\mu\}$, and
that generates zero norm states: ${\cal H}^\prime_o \doteq \Im^\prime_o \Omega
 \subset {\cal H}^\prime$. The 
field $ \mbox{\large{$\ell$}} = ( \zeta + \xi_b ) $, that acts as 
potential for the longitudinal 
current $\mbox{\large{$\ell$}}_\mu$, does not belong to the field 
algebra $\Im^\prime$ and only its space-time
derivatives occur in $\Im^\prime$. In this way, the exponential field $\Upsilon$ 
given by (\ref{spurious}) also does not belong to $\Im^\prime$ and
the Hilbert space cannot be factorized, ${\cal H}^\prime 
\neq {\cal H}_{_{\Upsilon}} \otimes {\cal H}_{_{\Psi}} $.

>From the algebraic point of view, the fact that the field $\Upsilon$ does
not belong to the field algebra $\Im^\prime$ and thus cannot be defined as an 
operator in ${\cal H}^\prime$, follows from the charge content of
${\cal H}_{_B}$ and ${\cal H}^\prime$, since some {\it topological} 
charges get trivialized in going from ${\cal H}^{^{B}}$ to ${\cal H}$
\cite{M}. One can define  conserved currents $j_\mu$ belonging to the Bose
field algebra $\Im^{^B}$, as for example,

\be
j^\mu (x) \doteq \frac{\beta}{2 \pi}\,
\epsilon_{\mu \nu}\,\partial^\nu \Phi (x) +  \partial^\mu \zeta\,\in\,
\Im^{^B}\,.
\ee

\noindent  The corresponding charges are

\be
{\cal Q} \doteq   \int_{- \infty}^{+ \infty} \,d z^1\,j^0 (z)\,,
\ee

\noindent such that

\be
\big [ {\cal Q} ,  \Im^{^{\!B}} \big ] \neq  0\,.
\ee

\noindent This implies that the charges ${\cal Q}$ do not vanish 
on $ {\cal H}^{^{\!B}}$:

\be
{\cal Q}\, {\cal H}^{^{\!B}} \neq 0\,.
\ee

\noindent The charges ${\cal Q}$ commute with $\psi$, $J^\mu$
and $\mbox{\large{$\ell$}}_\mu$, that is,

\be
\big [ {\cal Q} ,  \Im_o \big ] = 0\,\,\,,\,\,\, 
\big [ {\cal Q} ,  \Im^\prime \big ] = 0\,\,
\rightarrow \,\,\big [ {\cal Q} ,  \Im \big ] = 0\,.
\ee

\noindent This means that the charges ${\cal Q}$ are trivialized in the restriction
from $ {\cal H}^{^{\!B}}$ to ${\cal H}^\prime$ or ${\cal H}$ \cite{M,Bel,BRR}:

\be
{\cal Q}\,{\cal H}^{^{\!B}} \neq 0\,\,\,,\,\,\,
{\cal Q}\,{\cal H}^\prime = 0\,\,\,,\,\,\,
{\cal Q}\,{\cal H} = 0\,.
\ee

\noindent Since $\big [ {\cal Q} , \Upsilon \big ] = \alpha  \Upsilon
$, the state $  \Upsilon \Omega$ cannot
belong to ${\cal H}^\prime$ and the field $\Upsilon$ cannot be defined
as an operator in the Hilbert 
space ${\cal H}^\prime$ \cite{M,Bel,BRR}. Nevertheless, the 
spurious ``bilocal''operator $W_{_{\!b}} (x,y)$ is
neutral under ${\cal Q}$ and can be defined as an element of ${\cal H}^\prime$ and leads to 
constant correlation functions

\be
\langle \Omega \vert\, W_{_{\!b}} (x_1,y_1)\, W_{_{\!b}} (x_2,y_2) \cdots
 W_{_{\!b}} (x_n,y_n)\, \vert \Omega \rangle\, =\, 1\,.
\ee

\noindent The infinitely delocalized state $W_{_{\!b}} (x,y)\, \Omega$ 
is translationally invariant in ${\cal H}^\prime$. The position
independence of this state can be seen by computing the general
Wightman functions involving the  operator $W_{_{\!b}} (x,y)$ and all 
operators belonging to the local field algebra $\Im^\prime$. For any operator 

\be
{\cal O} (f_{_{\!z}}) = \int\, {\cal O} (z) f (z) d^2 z \,\in\,\Im^\prime\,,
\ee

\noindent of polynomials in the smeared fields defining the field algebra
$\Im^\prime$, the position independence of the operator
(\ref{w}) can be expressed in the weak form as

\be
\langle \Omega \vert \prod_{i = 1}^m
:e^{\,i\,g\,\int_{x_i}^{y_i} \,\ell_\mu (z) dz^\mu}:\,{\cal O}
(f_{_{\!z_1}}, \dots , f_{_{\!z_n}} ) \vert \Omega \rangle\,
 =\,F (z_1, \dots ,z_n) \equiv \langle \Omega \vert {\cal O} (f_{_{\!z_1}},
\dots ,f_{_{\!z_n}}) \vert \Omega \rangle\,,
\ee

\noindent where $ F (z_1, \dots ,z_n) $ is a distribution independent of the space-time coordinates
$(x_i,y_i)$. Since the operator $W_{_{\!b}} (x,y)$ carries no selection rule, it
reduces to the identity in ${\cal H}^\prime$. By the other hand, the operator
$\hat \sigma (x)$ does not reduces to the identity in ${\cal H}$ since it carries
the fermion chirality. As is well known, this operator gives rises to 
the existence of an infinite degeneracy of the ground-state implying the
$\theta$-vacuum parametrization \cite{LS,RS}.

The current $\ell_\mu$ commutes with itself and with 
all operators belonging to the field algebra $\Im^\prime$,

\be
[ \ell_\mu (x) , \Im^\prime ] = 0\,,
\ee

\noindent implying that,

\be
\langle \Theta^\prime \vert \ell_\mu (x) \vert \Xi^\prime \rangle = 0 =
\langle \Theta^\prime \vert\Big ( B_\mu (x) - g \bar \psi (x) \gamma_\mu
\psi (x) \Big ) \vert \Xi^\prime \rangle \,\,\,,\,\,\forall\,
\vert \Theta^\prime \rangle, \vert \Xi^\prime \rangle \in {\cal H}^\prime\,.
\ee

\noindent The states in $ {\cal H}^\prime$ can be
accomodated as equivalence classes modulo $\mbox{\large{$\ell$}}_\mu$, in such a way that 
the Hilbert space ${\cal H}$ is the quocient space 

\be
{\cal H} = \frac{ {\cal H}^\prime} {{\cal H}^\prime_o}
\,.
\ee

>From the operator point of view, the equivalence
established by Eq. (\ref{iso})  implies the 
algebraic isomorphism

\be
\Im\{\bar \psi,\psi,{\cal A}_\mu\} \sim \Im^{\prime\prime} 
\{\overline{\Psi}\Upsilon^\ast, \Psi \Upsilon,{\cal A}_\mu\} 
\sim \Im \{\overline{\Psi}, \Psi,{\cal A}_\mu\}\,,
\ee

\noindent where $ \Im^{\prime\prime} 
\{\overline{\Psi} \Upsilon^\ast, \Psi \Upsilon,{\cal A}_\mu\} \subset
\Im^{\prime} \{\Im^\prime_o,\overline{\Psi} \Upsilon^\ast, \Psi
\Upsilon, {\cal A}_\mu\}$. Since in the quocient space ${\cal H}$, the 
current $\ell_\mu$ is
the null element, $\ell \equiv 0$, within the functional integral formalism the
quocient space is obtained with $\rho_\mu = 0$. Taking into account the fact that the 
operator $W_{_{\!b}} (x,y)$ becomes the identity in ${\cal H}$,  we 
obtain the equivalence in the functional integral approach

\be
{\Z}^\prime [\upsilon,\bar \upsilon,\omega_\mu,0] \sim {\Z}
[\upsilon,\bar \upsilon,\omega_\mu ] 
\sim {\Z}_{_{\!{\Phi,\Sigma,\eta_a}}} [\upsilon,\bar \upsilon,
\omega_\mu ]\,,
\ee

\noindent where

\be
{\Z}_{_{\!{\Phi,\Sigma,\eta_a}}} [\upsilon,\bar \upsilon,
\omega_\mu ] =
\int\,{\cal D}  \Phi \,{\cal D} \eta_a\,{\cal D}  \Sigma\,
e^{\,\textstyle i\, S\, [ \Sigma,
\Phi, \eta_a]}\, e^{\,\textstyle 
i \int\,d^2x\,  \{  \overline{\Psi} \,\upsilon + \bar \upsilon\, \Psi 
\,+\,\frac{1}{m_\ast}\,\omega^\mu \epsilon_{\mu \nu} 
\partial ^\nu ( \Sigma -  \eta_a )\,\}}\,,
\ee

\noindent is the generating functional of the Schwinger-Thirring model with massive
fermions. This establishes the fermion-boson mapping 
between the massive Schwinger-Thirring 
model and 
the coupled sine-Gordon
theories in the positive semi-definite Hilbert space
${\cal H}$. 

The states in ${\cal H}$ can be accomodated as equivalent classes modulo $L_\mu$, in
such a way that the positive-definite gauge invariant 
Hilbert space $\hat {\cal H}$ is the quocient space

\be
\hat {\cal H} = \frac{{\cal H}}{{\cal H}_o}\,,
\ee

\noindent where ${\cal H}_o = L_\mu\,\Omega$.

The appearance in the mass term of the spurious 
field $\hat \sigma (x)$, that carry the free fermion field
chirality, signalize the explicit breakdown of the
chiral symmetry. For $M_o \rightarrow 0$, the fields $\Phi (x)$, $ \eta_a (x)$ becomes 
free and massless, $\Sigma$ is a massive free field, and we recover 
the massless Schwinger-Thirring model.

%%%%%%%%%%%%%%%%%%%%%%%%%%%%%%%%%%%%%%%%%%%%%%%%%%%%%
\section{Anomalous $QED_2$ with massive fermions}
%%%%%%%%%%%%%%%%%%%%%%%%%%%%%%%%%%%%%%%%%%%%%%%%%%%%%
\setcounter{equation}{0}

In this section we shall consider the functional integral bosonization of the 
anomalous $QED_2$ with massive Fermi fields, defined from the Lagrangian 
density

\be\label{csl}
{\cal L} = - \frac{1}{4} {\cal F}^2_{\mu \nu} + 
\chi^{\dagger}_{_{\!r}} i \partial_+ \chi_{_{\!r}}\,+\,
\psi^\dagger_{_{\!\ell}} ( i \partial_- - e {\cal A}_- ) \psi_{_{\!\ell}} - 
M_o ( \chi{^\dagger}_{_{\!r}} \psi_{_{\!\ell}} + 
\psi_{_{\!\ell}}^\dagger \chi_{_{\!r}})\,.
\ee

\noindent The anomalous model with massless Fermi fields was considered in
detail in Refs.\cite{Bel,Boy} using the operator approach. The so-called 
``decoupled'' massless free scalar Bose fields play an important role in the
bosonized massless chiral model 
in order
to ensure the existence of Fermi fields in the asymptotic states, the 
cluster decomposition property of the Wightman functions, the 
inexistence of $\theta$-vacuum and the isomorphism
between the gauge invariant and gauge non-invariant 
formulations \cite{BRR,BR}.  In what 
follows we shall discuss the role played by these fields in the corresponding
anomalous chiral  model with massive fermions.

%%%%%%%%%%%%%%%%%%%%%%%%%%%%%%%%%%%%%%%%%%%
\subsection{Functional integral bosonization}
%%%%%%%%%%%%%%%%%%%%%%%%%%%%%%%%%%%%%%%%%%%%

Following the same procedure of the previous section, we 
parametrize the gauge field in terms of the Bose fields $\{U,V\}$,

\be
{\cal A}_+ = - \frac{1}{e}\,U^{-1} i \partial_+ U\,,
\ee

\be
{\cal A}_- = - \frac{1}{e}\,V i \partial_- V^{-1}\,.
\ee

\noindent Introducing the decomposition for the gauge field \footnote{With 
the notation
of the preceding section, $\phi \equiv \phi_a$, $\xi \equiv \xi_a$.}

\be
{\cal A}_\mu =  \frac{\sqrt \pi}{e}\,\{ \epsilon_{\mu \nu} \partial^\nu
 \phi + \partial _\mu \xi \}\,,
\ee

\noindent we can write,

\be
U = e^{\textstyle\,i \sqrt \pi ( \phi + \xi)}\,,
\ee

\be
V = e^{\textstyle\,i \sqrt \pi ( \phi - \xi)}\,.
\ee

Performing the chiral fermion rotation

\be
\psi_{_{\!\ell}} = V \chi_{_{\!\ell}}\,,
\ee

\noindent the Jacobian associated with the change in the functional integral 
measure is given by

\be\label{j}
{\cal J} = e^{\textstyle\,- i \Gamma [V] + i \frac{a}{8\pi}\,\int d^2
z (U^{-1} i \partial_+ U )(V i \partial_- V^{-1})}\,.
\ee

\noindent The presence of the last factor in (\ref{j}) reflects the usual 
regularization  ambiguity, with $a$ the Jackiw-Rajaraman 
parameter \cite{Bel,BRR,Boy,JR}.

Defining the gauge invariant field $G \doteq UV$,  we can 
write the Jacobian  (\ref{j}) as

\be
{\cal J} = e^{\textstyle\,- i \frac{a}{2} \Gamma [G] +
i (\frac{a}{2} - 1 ) \Gamma [V] + i \frac{a}{2} \Gamma [U]}\,.
\ee

\noindent The total effective action is given by

$$
S _{_{eff.}} =  S _{_{\!M}} [G] -  
\frac{a}{2} \Gamma [G] + (\frac{a}{2} - 1 ) \Gamma [V] +  
\frac{a}{2} \Gamma [U] +
$$

\be\label{aa}
+ \int\,d^2 z\,\Big \{\,\bar \chi\,i \slash\!\!\!\partial\,\chi - M_o\,(\,
\chi^\ast_{_{\!r}}\chi_{_{\!\ell}}\,V^{-1}\,+\,
\chi^\ast_{_{\!\ell}}\chi_{_{\!r}}\,V \,\Big \}\,,
\ee

\noindent where $S_{_{\!M}} [G]$ is the Maxwell action,

\be
S_{_{\!M}} [G] = \frac{1}{8 e^2}\,\int\,d^2z\,\Big [ \partial_+\,\Big (\,
G\,i\,\partial_- G^{-1} \Big ) \Big ]^2\,.
\ee

\noindent For $a = 2$ and $M_o = 0$, the field $V$ decouples 
from (\ref{aa}) and 
except by the presence of the action $\Gamma [U]$, which
appears to play merely a spectator role, the effective action (\ref{aa}) 
is  that of the $QED_2$ with massless Fermi fields,

\be
S _{_{eff.}}\Big \vert_{_{\!{a=0}\atop{M =0}}}
= S _{_{Sch}} + \Gamma [U]\,.
\ee

\noindent As shown 
in Refs. \cite{Bel,BRR}, the apparently decoupled field $U$ plays an 
important role in the construction of the Hilbert space of the anomalous 
chiral model with massless fermions.

Following the same procedure of the preceding section, performing the 
bosonization of the free massive Fermi field $\chi$ in terms
of the Bose field $ \varphi$, the effective Lagrangian density can be 
written as

$$
{\cal L} = \frac{4 \pi}{2 e^2} (\Box  \phi )^2 + \frac{1}{2} (a + 1)
 \phi \Box  \phi -
\frac{1}{2} \xi^\prime \Box \xi^\prime - (a - 1)^{- 1/2} 
(\Box  \phi )\xi^\prime +
$$

\be\label{al}
- \frac{1}{2}  \varphi \Box  \varphi - M_o^\prime\,
 \cos \{ 2 \sqrt \pi  \varphi
- 2 \sqrt \pi ( \phi - (a - 1)^{-1/2}\xi^\prime )\}\,,
\ee

\noindent where we have defined the canonical field 

\be
\xi^\prime = (a - 1)^{1/2} \xi\,.
\ee

\noindent The effect of the anomaly introduces the 
field $\xi^\prime$, which is a pure gauge 
excitation in the vector model, in the effective
bosonized chiral model \cite{Bel,BRR,Boy}. Due to the presence of a 
mass term for the Fermi 
fields, the field $\xi^\prime$ no longer is a free field. As we shall see, the 
field $\xi^\prime$ plays an
important role in the field algebra and in the construction of the Hilbert
space of the anomalous chiral model with massive fermions.

The fields $\xi^\prime$ and 
$ \phi$ can be decoupled in (\ref{al}) using that

\be
- \frac{1}{2} \Big ( \xi^\prime \Box \xi^\prime + 2 (a -
1)^{- 1/2} (\Box  \phi ) \xi^\prime \Big ) =
 - \frac{1}{2} \xi^{\prime \prime} \Box \xi^{\prime\prime} +
\frac{1}{2} \phi^\prime \Box  \phi^\prime\,,
\ee

\noindent where

\be
\phi^\prime = \frac{a}{\sqrt{a-1}}\, \phi\,\,\,,\,\,\,
\xi^{\prime\prime} =  \xi^\prime +  \phi^\prime\,.
\ee

The effective Lagrangian density is given by

$$
{\cal L} =  \frac{1}{2 m_a^2} (\Box \phi^\prime )^2 +
 \frac{1}{2} \phi^\prime \Box \phi^\prime - 
\frac{1}{2} \xi^{\prime \prime} \Box \xi^{\prime
\prime} - \frac{1}{2} \varphi \Box \varphi - 
$$

\be\label{l}
- M_o \cos \Big \{ 2 \sqrt \pi
\varphi + 2 \sqrt{\frac{\pi}{a-1}} ( \xi^{\prime\prime} -
\phi^\prime ) \Big \}\,,
\ee

\noindent where 

\be
m_a^2 = (e\,a)^2\,/\,4 \pi (a-1)\,.
\ee

In a similar way, the higher-derivative term in (\ref{l}) can be 
reduced using that

$$
\int\,{\cal D} \,\phi^\prime\,
e^{\textstyle \, i \int d^2 z \Big \{ \frac{1}{2 m_a^2} (\Box
\phi^\prime )^2 + \frac{1}{2} \phi^\prime \Box \phi^\prime \Big \}}
\equiv 
$$

\be 
\int {\cal D}  \Sigma\,\int\,{\cal D}\,\phi^{\prime \prime}\,
\,e^{\textstyle \,i \int d^2 z \Big \{ -
\frac{1}{2}  \Sigma \Box  \Sigma - 
\frac{1}{2}m_a^2  \Sigma ^2 +  
\phi^{\prime\prime} \Box \phi^{\prime \prime} \Big \} } \,.
\ee

For further convenience we relabel the field $\phi^{\prime \prime} = \eta$, and
the effective bosonized Lagrangian density is then given by

$$
{\cal L} = \frac{1}{2} (\partial_\mu \xi^{\prime\prime} )^2
-
\frac{1}{2} (\partial_\mu \eta )^2 +
\frac{1}{2} (\partial_\mu \varphi)^2 +
\frac{1}{2} (\partial_\mu  \Sigma)^2 -
\frac{1}{2} m_a^2 ( \Sigma )^2-
$$

\be\label{ecl}
- M^\prime_o : \cos \Big \{ \,2 \sqrt{\frac{\pi}{a-1}}\, \Sigma\,+\,
 2 \sqrt \pi  \varphi\,+\,2 \sqrt{\frac{\pi}{a-1}}\,(\,
 \xi^{\prime\prime} - \eta\,)\,\Big \} :\,.
\ee

\noindent The mass term carry dependence  on the
fields $\xi^{\prime\prime}$ and
$\eta$, which are quantized with opposite metric, the massive 
field $\Sigma$ and
the field $\varphi$, which carries the free fermion degrees of 
freedom. Since the 
fields $\xi^{\prime\prime}$ and $\eta$ are quantized with opposite metric, the 
corresponding equations of motion are given by

\be
\Box \xi^{\prime\prime} \,-\,2 \Big ( \frac{\pi}{a-1} \Big )^{1/2}\,
 M^\prime_o : \sin \Big \{ \,2 \sqrt{\frac{\pi}{a-1}}\, \Sigma\,+\,
 2 \sqrt \pi  \varphi\,+\,2 \sqrt{\frac{\pi}{a-1}}\,(\,
 \xi^{\prime\prime} - \eta\,)\,\Big \} :\,=0\,,
\ee

\be
-\,\Box \eta \,+\,2 \Big ( \frac{\pi}{a-1} \Big )^{1/2}\,
 M^\prime_o : \sin \Big \{ \,2 \sqrt{\frac{\pi}{a-1}}\, \Sigma\,+\,
 2 \sqrt \pi  \varphi\,+\,2 \sqrt{\frac{\pi}{a-1}}\,(\,
 \xi^{\prime\prime} - \eta\,)\,\Big \} :\,=0\,.
\ee

\noindent Although the fields $\xi^{\prime\prime}$ and $\eta$ acquire a
non trivial dynamics,  the field 
combination $ ( \xi^{\prime\prime}\,-\,\eta ) $ is a free and 
massless field 

\be
\Box ( \xi^{\prime\prime}\,-\,\eta ) = 0\,,
\ee

\noindent that generates zero norm states from the 
vacuum. For massless fermions ($M_o = 0$) the 
fields $\xi^{\prime\prime}$ and $\eta$, becomes massless free fields 
and even in this case we cannot disregard them from the 
field algebra \cite{Bel}.

%%%%%%%%%%%%%%%%%%%%%%%%%%%%%%%%%%%%%%%%%%%
\subsection{Field algebra}
%%%%%%%%%%%%%%%%%%%%%%%%%%%%%%%%%%%%%%

The set of 
fields $\{\bar \psi,\psi,{\cal A}_\mu\}$
defines the field algebra $\Im$ and constitute the 
intrinsic mathematical description of the model. In terms of the 
Bose fields, the Fermi field  can be written as

\be
\psi (x)\, =\,: e^{\,\textstyle i \,\sqrt \pi\,
 \frac{(1 + \gamma^5)}{2}\,[\phi(x) - \xi (x)]}:\,
\chi (x)\, =\, \Psi (x)\,\omega (x)\,,
\ee

\noindent where

\be
\Psi (x)\,=\,
:e^{\,\textstyle i (1 + \gamma^5)\, \sqrt{\frac{\pi}{a-1}}\,  \Sigma (x)}:\,
\chi (x)\,,
\ee

\be
\omega (x) = : e^{\,\textstyle i\,(1 + \gamma^5)\, \sqrt{\frac{\pi}{a-1}}\,
(\xi^{\prime\prime} - \eta )}:\,,
\ee

\be
\chi (x) = \Big ( \frac{\mu_o}{2 \pi} \Big )^{1/2}\,
e^{\textstyle \, - i \frac{\pi}{4}\,\gamma^5}\,
:e^{{\textstyle
i\,\sqrt \pi \{\,\gamma^5\, \varphi (x) + }\int_{x^1}^{ \infty}\,
\textstyle {\dot{ \varphi}} (x^0,z^1)\,d z^1 \}}:\,.
\ee

The gauge field can be written as

\be
{\cal A}_\mu (x)\, =\,\Sigma_\mu (x)\,+\,\frac{1}{e}\,\omega^{-1} (x)\,\partial_\mu\,\omega (x)\,.
\ee

\noindent where

\be
\Sigma_\mu (x)\,\doteq\, - \frac{1}{m_{_{\!a}}}\,\Big \{\,
\epsilon_{\mu \nu}\, \partial^\nu\,+\,
\frac{1}{a-1}\,\partial_\mu\,\Big \}\,\Sigma (x)\,.
\ee

The vector current is given by

\be\label{Jv}
J_\mu = m_a\,\epsilon_{\mu \nu} \partial^\nu\,\Sigma\,+\,L^\mu\,,
\ee

\noindent where $L_\mu$ is a longitudinal current\,

\be
L_\mu = - \frac{1}{\sqrt \pi}\,
\Big \{\,(\partial_\mu\,+\,\tilde \partial_\mu ) \varphi\,+\,
\frac{a}{\sqrt{a-1}}\,\tilde \partial_\mu \eta\,-\,\frac{1}{\sqrt{a-1}}\, [\,
(a-1)\,\partial_\mu\,-\,\tilde \partial_\mu ] \xi^{\prime\prime} \Big \}\,.
\ee

\noindent For $M_o = 0$ we recover the operator solution of the anomalous model
with massless fermions discussed in Refs. \cite{Bel,Boy}.

Due to the opposite metric quantization for the coupled sine-Gordon 
fields $\xi^{\prime\prime}$ and $\eta$, the massless free field 
combination $ (\xi^{\prime\prime} -\eta)$ generates zero norm 
states from the vacuum

\be
\Vert \big ( \xi^{\prime\prime} (x)  - \eta (x) \big ) 
\vert \Omega \rangle \Vert^2 = 0\,.
\ee

\noindent As a consequence, the field $\omega (x)$ generates constant 
contributions to the Wightman functions

\be
\langle \Omega \vert \prod_{i=1}^n \omega^\ast (x_i)\,\prod_{j=1}^n \omega (y_j)
\vert \Omega \rangle = 1\,.
\ee

Due to the presence of the longitudinal current $L_\mu$ in Eq. (\ref{Jv}), the 
Gauss' law holds in the weak form and is satisfied on the physical 
subspace ${\cal H}^{^{\!phys}}$, which is  defined by the subsidiary condition

\be
\langle \Phi \vert ( \partial_\mu\,{\cal F}^{\mu \nu}(x) + \,J^\nu(x))\vert
\Psi \rangle\,=\,\langle \Phi \vert L^\nu(x) \vert \Psi \rangle\,\,\,\,,\,\,\,
\forall \vert \Phi \rangle, \vert \Psi \rangle \in {\cal H}^{^{\!phys}}\,.
\ee

In a genuine gauge theory, the algebra of physical 
operators $\Im^{^{\!phys}}$ must be identified
with the subalgebra of $\Im$ which obeys the subsidiary condition and 
the physical
Hilbert space is defined by gauge invariant states accomodated as equivalent 
classes modulo $L_\mu (x)$. However, it is a peculiarity of two-dimensional 
anomalous chiral models \cite{Bel,BRR} that the 
algebra $\Im^{^{\!phys}} \equiv \Im$, since
all operators belonging to the intrinsic field algebra $\Im$ commute with
the longitudinal current, 

\be
[ {\cal O} , L_\mu ] = 0\,\,\,,\,\,\,\forall {\cal O} \in \Im\,.
\ee

\noindent The chiral anomaly promotes the intrinsic 
set of fields $\{\psi,\psi,{\cal A}_\mu\}$ to the status of physical 
operators, which must be singlet under chiral operator gauge 
transformations. We shall return to this point further on.

It is very instuctive to compare the mass term of the anomalous 
model, appearing 
in (\ref{ecl}) with that of the gauge model (\ref{sg}) with $g = 0$. For the
Schwinger model we 
have $\beta = 2 \sqrt \pi$,\,
$\Phi \equiv \varphi$, $\eta_a \equiv \eta = \phi^{\prime\prime}$, and
the mass operator is given by

\be\label{gmo}
:\cos \Big \{\,2 \sqrt \pi \,\Sigma  + 2 \sqrt \pi ( \varphi  - \eta  ) \Big \}:\,.
\ee

\noindent  The spurious field combination 

$$ 
\hat \sigma  = :e^{\,2 i \sqrt \pi (\varphi  - \eta )}:\,,
$$

\noindent carries the free fermion 
selection rule and the violation of the asymptotic factorization property 
of the Wightman functions leads
to the $\theta$-vacuum structure \cite{LS,RS,B,AAR}. 

In the anomalous model, the spurious field
combination $( \xi^{\prime\prime} - \eta )$ carries no fermion selection 
rule, which
leads to the absence of vacuum degeneracy. For $a = 2$ the 
mass operator of the  anomalous model can be written as

\be\label{amo}
:\cos \Big \{\,2 \sqrt \pi \,\Sigma + 2 \sqrt \pi \varphi + 
2 \sqrt \pi\,( \xi^{\prime \prime} - \eta ) \Big \}:\,.
\ee

\noindent The presence of the field $\xi^{\prime\prime}$ in the mass operator
(\ref{amo}) explicitly shows that for $a = 2$ the anomalous chiral model 
with massive fermions cannot be considered as equivalent to the massive
$QED_2$, as has been 
claimed \cite{Carena} for the corresponding models with massless 
fermions.  For $M_o = 0$, the Bose fields $\xi^{\prime\prime}$ and
$\eta$ becomes free massless fields. Although in this case we can 
write the chiral density operator as

\be 
[\psi^\ast_{_{\!\ell}}\chi_{_{\!r}}]\,
=\, :e^{\,2 i \sqrt \pi \Sigma}: \hat \sigma\,
:e^{\,2 i \sqrt \pi \xi^{\prime\prime}}:\,,
\ee

\noindent the field $\xi^{\prime\prime}$ cannot be extracted  from
the field algebra and the field $\hat \sigma$ cannot be defined
on the Hilbert space of the anomalous chiral model \cite{Bel,BRR}.

In the gauge model previously considered, the contributions of the 
field $\eta$, quantized with negative metric, cancels those arising from the
field $\varphi$ and which carries the fermion degrees of freedom. As is well known 
\cite {LS,RS,S,B}, this charge-screening mechanism leads to the 
vacuum degeneracy and
the $\theta$-vacuum parametrization. However, in the anomalous model, the 
appearance in the mass operator of the massless free field 
combination $(\xi^{\prime\prime} - \eta )$, has no physical 
consequences on the vacuum structure of the model. This can be shown by
considering for example the Wightman functions of the mass 
operator $[ \bar \psi (x) \psi (x) ]$. Using the functional integral 
Gell-Mann and Low formula, the functional integration over the field $\eta$ 
cancels those arising from the integration over the field $\xi^{\prime\prime}$ 
and we get, 

$$
\langle \Omega \vert [\bar \psi (x_1) \psi (x_1)] \cdots
[ \bar \psi (x_n) \psi (x_n)] \vert \Omega \rangle =
$$

$$
\langle \Omega \vert : \cos \Big \{ \,2 \sqrt{\frac{\pi}{a-1}}\, \Sigma (x_1)
\,+\, 2 \sqrt \pi  \varphi (x_1)\,+\,2 \sqrt{\frac{\pi}{a-1}}\,(\,
\xi^{\prime\prime}(x_1) - \eta (x_1)\,)\,\Big \} : \cdots
$$

$$
: \cos \Big \{ \,2  \sqrt{\frac{\pi}{a-1}}\, \Sigma (x_n)
\,+\, 2 \sqrt \pi  \varphi (x_n)\,+\,2 \sqrt{\frac{\pi}{a-1}}\,(\,
\xi^{\prime\prime}(x_n) - \eta (x_n)\,)\,\Big \} : \vert \Omega \rangle \equiv
$$

$$
\langle \Omega \vert : \cos \Big \{ \,2  \sqrt{\frac{\pi}{a-1}}\, \Sigma (x_1)
\,+\, 2 \sqrt \pi  \varphi (x_1) \Big \} : \cdots
: \cos \Big \{ \,2 \sqrt{\frac{\pi}{a-1}}\, \Sigma (x_n)
\,+\, 2 \sqrt \pi  \varphi (x_n) \Big \} : \vert \Omega \rangle =
$$

\be
= \langle \Omega \vert [\overline \Psi (x_1) \Psi (x_1)] \cdots
[\overline \Psi (x_n) \Psi (x_n)] \vert \Omega \rangle \,.
\ee

\noindent For the general Wightman functions obtained 
with polynomials of the fundamental
fields defining the model,  we obtain the isomorphism 

\be\label{isowf}
\langle \Omega \vert {\cal P} (\bar \psi,\psi,{\cal A}_\mu) \vert \Omega \rangle 
\equiv
\langle \Omega \vert {\cal P} (\overline \Psi, \Psi,\Sigma_\mu) \vert \Omega \rangle \,,
\ee

\noindent implying that the quantum dynamics and the net 
physical content of the model are carried by the
fields $\{\bar \Psi,\Psi,\Sigma_\mu\}$.

%%%%%%%%%%%%%%%%%%%%%%%%%%%%%%%%%%%%%%%%%%%%%%%%%%%%%%%%%%
\subsection{Wess-Zumino field and extended gauge invariance}
%%%%%%%%%%%%%%%%%%%%%%%%%%%%%%%%%%%%%%%%%%%%%%%%%%%%%%%%%%

Although the Bose fields $\xi^\prime$, $\eta$ and $\omega$ gives no 
contributions to the general Wightman functions of the anomalous chiral 
model, they play an important role in the field algebra and in the
structure of the  Hilbert space. In this section 
we shall display the role played by these fields and the 
Wess-Zumino (W-Z) field in order to
ensure the invariance of the field algebra under extended 
local gauge transformations.

To begin with, consider the action of the anomalous model in terms of the
field variables $\{U,V\}$,

$$
S [U,V,\bar \chi,\chi ]  = S [UV] + \Big ( \frac{a}{2} - 1 \Big ) \Gamma [V] +
\frac{a}{2} \Gamma [U] + 
$$

\be\label{W}
+ \bar \chi\, i\,
 \slash\!\!\!\partial\,\chi
- M_o \int d^2 z \Big ( \chi^\ast_{_{\!r}} \chi_{_{\!\ell}} V^{-1} +
\chi^\ast_{_{\!\ell}} \chi_{_{\!r}} V \Big )\,.
\ee

\noindent The $U (1)$ group-valued Bose fields $\{U,V\}$ can be factorized as,

\be
U = \bar U\,h\,\,\,,\,\,\,V = h^{-1} \bar V\,,
\ee

\noindent where $\{\bar U,\bar V\}$ depends on the 
field $\phi^\prime = \Sigma - \eta$,

\be 
\bar U = e^{\,\textstyle - 2 i\,\frac{(a - 2)}{a}\,\sqrt{\frac{\pi}{a-1}}\,
\phi^\prime}\,,
\ee

\be 
\bar V = e^{\,\textstyle - 2 i\,\sqrt{\frac{\pi}{a-1}}\,\phi^\prime}\,,
\ee

\noindent  and the field variable $h$ carries the dependence on the 
field $\xi^{\prime\prime}$,

\be
h = e^{\,\textstyle 2 i \sqrt{\frac{\pi}{a-1}}\,\xi^{\prime\prime}}\,.
\ee

\noindent  The action (\ref{W}) can be written in terms of the 
fields $\{\bar U, \bar V, h\}$. Using the P-W identity (\ref{PW}), and the
fact that 

\be
( a - 2 ) \bar V \partial _\mu \bar V^{-1} - 
 a  \bar U \partial_\mu \bar U^{-1} = 0\,,
\ee

\noindent we obtain,

$$
S [U, V,\bar \chi,\chi ] \equiv 
S [\bar U,\bar V,h,\bar \chi,\chi ]  = 
S [\bar U \bar V ] + (a - 1) \Gamma [h] + 
\Big ( \frac{a}{2} - 1 \Big ) \Gamma [\bar V] +
\frac{a}{2} \Gamma [\bar U] + 
$$

\be\label{abar}
+ \bar \chi\, i\,
 \slash\!\!\!\partial\,\chi
- M_o \int d^2 z \Big ( \chi^\ast_{_{\!r}} \chi_{_{\!\ell}} \bar V^{-1} h
 + \chi^\ast_{_{\!\ell}} \chi_{_{\!r}} h^{-1} \bar V  \Big )\,.
\ee
                
The ``embedded'' version of the model is obtained by the introduction of 
the W-Z field by performing the chiral gauge transformations \cite{BR}

\be
^{^{g}}\!\!\psi_{_{\!\ell}} (x) = g (x) \psi_{_{\!\ell}} (x)\,,
\ee

\be
^{^{g}}\!\!\!{\cal A}_\mu (x) = {\cal A}_\mu (x) + \frac{1}{e}\,
 g (x)\, i\, \partial_\mu\, g^{-1} (x)\,,
\ee

\noindent where $g (x)$ is given in terms of the W-Z field $\theta (x)$,

\be
g (x) = e^{\,\textstyle - i \,(1 + \gamma^5)\,\sqrt \pi \,\theta (x)}\,.
\ee

\noindent The gauge transformation acts on the  Bose fields $\{U,V\}$ as,

\be
^{^{g}}\!U = U g\,\,\,\,,\,\,\,\, ^{^{g}}\!V = g^{-1} V\,\,\,\,,\,\,\,\,
 ^{^{g}}\!G = G\,.
\ee

\noindent The fields $\{\,^{^{g}}\!\!\bar \psi_{_{\!\ell}},\,
^{^{g}}\!\!\psi_{_{\!\ell}},\, ^{^{g}}\!\!\!{\cal A}_\mu\}$ are manifest 
invariant under extended gauge transformations:

\be
U \rightarrow U \tilde g\,\,,\,\,V \rightarrow \tilde g^{-1} V\,\,,\,\,
g \rightarrow g \tilde g^{-1}\,.
\ee

Rescalling the  W-Z field, 

\be
\theta^\prime = (a-1)^{1/2} \theta\,\,\,,\,\,\,
g = e^{\,\textstyle -\,2 i \sqrt{\frac{\pi}{a-1}}\,\theta^\prime}\,,
\ee

\noindent the gauge transformed variables $\{\,^g U,\,^g V\}$ are given by

\be  
^g U = U g = \bar U h g \,,
\ee

\be  
^g V =  g^{-1} V  = (h g)^{-1} \bar V \,.
\ee

\noindent Introducing the field 

\be
\xi^{\prime \prime \prime} \doteq \xi^{\prime\prime} - \theta^\prime\,,
\ee

\noindent we can define a new field variable $\tilde h$,

\be
\tilde h = e^{\,\textstyle 2 i \sqrt{\frac{\pi}{a-1}}\,( \xi^{\prime\prime} - 
\theta^\prime)} = 
e^{\,\textstyle 2 i \sqrt{\frac{\pi}{a-1}}\,\xi^{\prime\prime\prime}}\,.
\ee

\noindent In this way, we obtain the algebraic isomorphism

\be\label{gtu}  
^g U =  \bar U \tilde h \sim \bar U h = U\,,
\ee

\be\label{gtv}  
^g V =  \tilde h^{-1} \bar V \sim  h^{-1} \bar V = V\,,
\ee

\noindent showing that, from the algebraic point of view the two 
descriptions are indeed equivalent. 

For the general Wightman functions of the field operators belonging
to the intrisic field algebra we obtain 

\be
\langle \Omega \vert\, ^{^{g}}\!\! \bar \psi (x_1) \cdots ^{^{g}}\!\! \bar \psi (x_n)
^{^{g}}\!\!\psi (y_1) \cdots ^{^{g}}\!\!\psi (y_n) \vert \Omega \rangle \equiv
\langle \Omega \vert  \bar \psi (x_1) \cdots  \bar \psi (x_n)
\psi (y_1) \cdots \psi (y_n) \vert \Omega \rangle\,,
\ee

\be
\langle \Omega \vert\, ^{^g}\!\!\!{\cal A}_\mu (x_1) \cdots ^{^g}\!\!\!{\cal A}_\mu (x_n)
\vert \Omega \rangle \equiv
\langle \Omega \vert {\cal A}_\mu (x_1) \cdots {\cal A}_\mu (x_n) \vert \Omega \rangle\,,
\ee

\noindent expressing the equivalence between the gauge transformed intrinsic
set of 
fields $\{\,^{^{g}}\!\!\bar \psi,\, ^{^{g}}\!\!\psi ,\,^{^g}\!\!\!{\cal A}_\mu\}$ 
and $\{\bar \psi, \psi, {\cal A}_\mu \}$:

\be\label{p}
^{^{g}}\!\!\psi (x)\, = \,\Psi (x)\,\tilde \omega (x)\,\sim\, \psi (x)\,,
\ee

\be\label{a}
^{^g}\!\!\!{\cal A}_\mu (x)\,=\,\Sigma_\mu (x)\,+\,\frac{1}{e}\,\tilde \omega^{-1} (x)\,
\partial_\mu\,\tilde \omega (x)\,\sim\,{\cal A}_\mu (x)\,,
\ee

\noindent where the field $ \tilde \omega (x)$ is written in 
terms of $\xi^{\prime\prime\prime}$,

\be
\tilde \omega (x) = : e^{\,\textstyle i\, 2\, \sqrt{\frac{\pi}{a-1}}\,
(\xi^{\prime\prime\prime} - \eta )}:\,.
\ee

\noindent This implies the isomorphism between the field algebras

\be
^{^g} \Im\,\sim\,\Im\,.
\ee

\noindent In this way, for any 
functional ${\opF} (\bar \psi, \psi, {\cal A}_\mu )$, we obtain 

\be
\langle \,\opF (\, ^{^g}\!\!\!{\cal A}_\mu,\, ^{^g}\!\!\bar \psi,\,
 ^{^g}\!\!\psi )\,
\rangle\,=\,
\langle \,\opF ( {\cal A}_\mu, \bar \psi, \psi )\,\rangle\,,
\ee

\noindent expressing the isomorphism of the Hilbert spaces:

\be
^{g} {\cal H} \sim {\cal H}\,.
\ee

In order to display the role played by the W-Z field in the model with massive
Fermi fields, consider the 
action corresponding to the model defined in terms of 
the gauge transformed fields $\{\,^{^{g}}\!\!\bar \psi_{_{\!\ell}},\,
^{^{g}}\!\!\psi_{_{\!\ell}},\, ^{^{g}}\!\!\!{\cal A}_\mu\}$, 

$$
S [Ug,g^{-1}V,\bar \chi,\chi ]= S [UV] + 
\Big ( \frac{a}{2} - 1 \Big ) \Gamma [g V] +
\frac{a}{2} \Gamma [U g^{-1}] + 
$$

\be \label{gta}
\bar \chi\, i\,
 \slash\!\!\!\partial\,\chi
- M_o \int d^2 z \Big ( \chi^\ast_{_{\!r}} \chi_{_{\!\ell}} V^{-1}g +
\chi^\ast_{_{\!\ell}} \chi_{_{\!r}} g^{-1} V \Big )\,.
\ee

\noindent Since we have performed an operator  chiral gauge transformation, the
mass term is not manifest invariant and the W-Z field obeys a coupled
sine-Gordon equation. Using the P-W identity, we 
can write the gauge transformed action (\ref{gta}) as

$$
S [Ug,g^{-1}V,\bar \chi,\chi ]=  S _{_{\!WZ}}[U,V,g] 
+ S [UV] + 
\Big ( \frac{a}{2} - 1 \Big ) \Gamma [V] +
\frac{a}{2} \Gamma [U] + 
$$

\be\label{gtawz} 
+ \bar \chi\, i\,
 \slash\!\!\!\partial\,\chi
- M_o \int d^2 z \Big ( \chi^\ast_{_{\!r}} \chi_{_{\!\ell}} V^{-1}g +
\chi^\ast_{_{\!\ell}} \chi_{_{\!r}} g^{-1} V \Big ) \,,
\ee
 
\noindent where $S _{_{\!WZ}}[U,V,g]$ is the W-Z action for the anomalous model
with massless Fermi fields ($M_o=0$) and is given by

\be\label{wza}
S _{_{\!WZ}}[U,V,g] = (a - 1) \Gamma [g] + \frac{1}{4 \pi}\,\int\,
d^2 z\,\Big \{\,
\Big ( \frac{a}{2} - 1 \Big ) V \partial^\mu V^{-1}\,-\,\frac{a}{2}\,U 
\partial^\mu U^{-1} \Big \}\,g \partial_\mu g^{-1}\,.
\ee

\noindent In terms of the factorized fields $\{\bar U,\bar V,h\}$, the 
WZ action can be written as

\be\label{w}
S _{_{\!WZ}}[U,V,g] \equiv S _{_{\!WZ}}[g,h]=
(a - 1) \Gamma [hg] - (a-1) \Gamma [g] - (a-1) \Gamma [h]\,.
\ee

\noindent From (\ref{gtawz}) and (\ref{w}) we 
obtain the gauge-transformed action as

$$
S [gU,g^{-1}V,\bar \chi,\chi ]  = 
S [\bar U, \bar V, \tilde h,\bar \chi,\chi] =
$$

\be
= S [\bar U \bar V] + (a - 1) \Gamma [\tilde h] +
\Big ( \frac{a}{2} - 1 \Big ) \Gamma [\bar V] +
\frac{a}{2} \Gamma [\bar U ] + \bar \chi\, i\,
 \slash\!\!\!\partial\,\chi
- M_o \int d^2 z \Big ( \chi^\ast_{_{\!r}} \chi_{_{\!\ell}} \bar V^{-1} \tilde h
 +
\chi^\ast_{_{\!\ell}} \chi_{_{\!r}} \tilde h^{-1} \bar V \Big )\,,
\ee

\noindent which is identical to the action (\ref{abar}) with 
the field $\xi^{\prime\prime}$ being replaced by 
the field $\xi^{\prime\prime\prime}$, and thus

\be\label{s}
S [Ug, g^{-1}V, \bar \chi, \chi ] \equiv
S [U,V,\bar \chi, \chi ]\,.
\ee

>From the functional integral point of view, the 
isomorphism of the Hilbert spaces $^{g} {\cal H}$ and  ${\cal H}$, follows 
from the equivalence between the corresponding generating functionals. As a 
consequence of
the algebraic isomorphism $^{^g} \Im \sim \Im$,  according 
to (\ref{p}) and (\ref{a}), the source terms 
associated with the intrinsic set of fields are singlets under 
extended gauge transformations. Using (\ref{gtu}), (\ref{gtv}),(\ref{s}), we 
obtain

$$
\langle\,e^{\,\textstyle i\,[\,^{^g}\!\!{\bar \psi}\, \upsilon\,+\,
\bar \upsilon\, ^{^g}\!\!\psi\,+\,^{^g}\!\!\!{\cal A}_\mu \,\rho^\mu\,]}\,
\rangle_{_{\!\Omega}}\,=\,
\langle\,e^{\,\textstyle i\,[\,\bar \Psi\, {\tilde \omega}^\ast\, \upsilon\,+\,
\bar \upsilon\, \Psi\, \tilde \omega\,+\,(\Sigma_\mu + \frac{1}{e}\,\tilde \omega^{-1}\,
\partial_\mu\,\tilde \omega\,) \rho^\mu\,]}\,\rangle_{_{\!\Omega}}\,
 \equiv\,
$$

\be
\langle\,e^{\,\textstyle i\,[\,\bar \Psi\,{\omega}^\ast\, \upsilon\,+\,
\bar \upsilon\, \Psi\,\omega\,+\,(\Sigma_\mu + \frac{1}{e}\,\omega^{-1}\,
\partial_\mu\,\omega\,) \rho^\mu\,]}\,\rangle_{_{\!\Omega}} \,
\equiv
\langle\,e^{\,\textstyle i\,[\,\bar \Psi\, \upsilon\,+\,
\bar \upsilon\, \Psi\,+\,\Sigma_\mu\,\rho^\mu\,]}\,\rangle_{_{\!\Omega}} \,,
\ee

\noindent which implies the equivalence between the generating functionals

\be
\langle\,e^{\,\textstyle i\,[\,^{^g}\!\!{\bar \psi}\, \upsilon\,+\,
\bar \upsilon\, ^{^g}\!\!\psi\,+\,^{^g}\!\!\!{\cal A}_\mu \,\rho^\mu\,]}\,
\rangle_{_{\!\Omega}}\,=\,
\langle\,e^{\,\textstyle i\,[\, \bar \psi\, \upsilon\,+\,
\bar \upsilon\, \psi\, +\, {\cal A}_\mu \,\rho^\mu\,]}\,
\rangle_{_{\!\Omega}}\,.
\ee

\noindent We conclude that, even in the anomalous chiral model with massive Fermi 
fields, the generating functional is invariant under extended gauge 
transformations. This implies the isomorphism between the Hilbert 
space ${\cal H}$ of the
anomalous model, defined in 
terms of the intrinsic 
fields $\{\bar \psi,\psi,{\cal A}_\mu\}$ and which are not manifest gauge 
invariant (gauge non-invariant formulation), and the 
Hilbert space $^g {\cal H}$ construct in terms
of the fields $\{\,^{^{g}}\!\!\bar \psi_{_{\!\ell}},\,
^{^{g}}\!\!\psi_{_{\!\ell}},\, ^{^{g}}\!\!\!{\cal A}_\mu\}$ and which are 
manifest invariant under extended gauge 
transformations (gauge invariant formulation). The role played by 
the W-Z field in 
the anomalous chiral model with massive fermions is exactly the same played
by it in the corresponding model with massless fermions. The introduction of 
the W-Z field in the anomalous chiral model replicates the theory, changing 
neither its algebraic structure nor its physical content. This
streamlines the conclusions of Refs. \cite{Bel,Boy} for the anomalous model
with massless Fermi fields.

%%%%%%%%%%%%%%%%%%%%%%%%%%%%%%%%%%%%%%%%%%%%%%%
\section{Conclusions}
%%%%%%%%%%%%%%%%%%%%%%%%%%%%%%%%%%%%%%%%%%%

Using a synthesis of the functional integral and operator formulations, we 
have considered some structural aspects of the fermion-boson mapping 
in two-dimensional gauge and anomalous gauge models. We have analyzed 
the role played by the ``decoupled'' free massless 
Bose fields, which appear in bosonized models with massless fermions, in 
the corresponding models with massive Fermi fields. 

For the $QED_2$ with current-current interaction among massive fermions, the 
use of an auxiliary vector
field to reduce the Thirring interaction, introduces a redundant Bose field
algebra which is insensitive to the presence of the mass term for the Fermi
fields. This procedure leads to the appearance of the longitudinal 
current $\mbox{\large$\ell$}_\mu$, which generates zero norm states 
from the vacuum.

For non Abelian models in $2 + 1$ dimensions, the 
Wilson loop operator \cite{Shapo,fradshapo} defined by,

$$
\exp\,\{g \int_\Gamma \bar \psi \gamma^\mu \psi dx_\mu \}\,,
$$

\noindent plays an important role in order to establish 
the fermion-boson correspondences. In 
two-dimensional Abelian 
models, due to the trivial topology concerned, in 
the computation of the expectation value of the loop operator,
 
$$
\langle\,\exp\,\{g \int_\Gamma \bar \psi \gamma^\mu \psi dx_\mu \} \rangle
= \langle\,\exp\,\{\int_\Gamma (B^\mu - \mbox{\large$\ell$}^\mu )
dx_\mu \} \rangle\,,
$$

\noindent the contributions of the current $\ell_\mu$ reduces to the 
identity. However, for non Abelian models, it seems 
to be very intructive to make a foundational 
investigation of the structural properties of the fermion-boson
mappings in $2 + 1$ dimensions, which may offer a valuable lesson for the
understanding of the underlying physical properties of the
higher dimensional field theory models. An interesting mathematical 
and structural question, that must 
be presumably 
relevant for the extension of the bosonization procedure to non-Abelian models
in $2 + 1$ dimensions \cite{Shapo}, where the knot invariants with non trivial
topology are present, is related to whether these loops operators belongs to 
the intrinsic field algebra and can be defined in the Hilbert space of the 
theory. Another question, that until recent publications has been not 
fully clarified, concerns to the appearance
of a local gauge symmetry in the bosonized version of the Thirring 
model in $2+1$ dimensions. A clear understanding of
these points seems to us essential in order to ensure that the
fermion-boson mapping is established on the Hilbert space of states and thus 
may offer information about the true physical content
of the original theory. 

In the anomalous chiral $QED_2$ with massive Fermi fields we show that the
original decoupled massless Bose fields quantized with opposite metric of the
chiral model with massless Fermi fields, are promoted to fields with 
non trivial dynamics  governed
by the sine-Gordon equation. Nevertheless, their combination remains
a free massless field with zero norm, which contributes with a phase to the
mass operator. Contrary to what happens in the genuine gauge model, this phase
factor carries no fermion selection rule and no vacuum degeneracy is implied by
the appearance of this spurious field with zero scale dimension. We have used
the introduction
of the mass term for the Fermi fields as an alternative and  practical way for
probing our previous conclusions of
Refs. \cite{Bel,BRR}. In that case, since  the Fermi field is massless, the 
decoupled massless Bose field appears at a first glance  to play merely a 
spectator role. However, the apparently decoupled Bose field plays an important
role in the construction of the Hilbert space of the anomalous chiral 
model. The naive extraction of this apparently decoupled massless free 
field from the field algebra, by factorizing the bosonized partition 
function, leads to misleading conclusions 
concerning to the physical content of the model, such as the violation of the 
asymptotic factorization property of the Wightman functions, the need of 
the $\theta$-vacuum parametrization and the equivalence of the chiral model 
defined for $a=2$ and the vector model, as proposed in Ref. \cite{Carena}. Even
in the anomalous model with massive Fermi fields, the introduction 
of the W-Z field only replicates the field algebra of the theory, changing
neither its algebraic structure not its physical content. As expected from the
analysis of the massless model \cite{Bel,BRR}, the Hilbert space of the 
anomalous chiral model with massive fermions and defined for $a = 2$ does 
not contains as a proper subspace the Hilbert space of the
corresponding gauge model.

{\small The authors are gratefull to CNPq-Brazil, FAPESP-Brazil and CAPES-Brazil
for partial financial
support. One of us (LVB) is gratefull to  R. L. P. G. Amaral and K. D. 
Rothe for many stimulating discussions.}
\\

%{\small{

%}}


\begin{thebibliography}{12}

%1
\bibitem{Shapo} M. Brali\'c, E. Fradkin, V. Manias, F. A. Shaposnik, Nucl. 
Phys.{\bf B 446} (1995) 144; E. Fradkin, F. A. Shaposnik, Phys. Lett. {\bf B 338} (1991) 243. 

%2
\bibitem{fradshapo} E. Fradkin and F. A. Shaposnik, Phys. Lett. 
B {\bf 338} 253 (1994);
S. Deser and R. Jackiw, Phys. Lett. B {\bf 139} 371 (1984). 


%3
\bibitem{FIB} L. V. Belvedere, Jour. Phys. {\bf A 33} (2000) 2755.

%4
\bibitem{M} G. Morchio, D. Pierotti and F. Strocchi, Ann. Phys.
(N.Y.) {\bf 188} (1988) 217; A. Z. Capri and R. Ferrari, Nuovo 
Cimento {\bf A 62} (1981) 273.  F. 
Strocchi, ``Selected Topics on the General Properties of 
Quantum Field Theory'', Lecture Notes in Physics, vol.51, World 
Scientific Publishing, 1993.

%5
\bibitem{Bel} C. G. Carvalhaes, L. V. Belvedere, H. Boschi Filho and
C. P. Natividade, Ann. Phys. {\bf 258} (1997) 210; C. G. Carvalhaes, L. V. Belvedere, R. L. P. G. do Amaral and N. A.
Lemos, Ann. Phys. {\bf 269} (1998) 1.

%6
\bibitem{BRR} R. L. P. G. do Amaral, L. V. Belvedere, K. D. Rothe and
F. G. Scholtz, Ann. Phys. {\bf 262} (1998) 132; L. V. Belvedere, R. L. P. G. Amaral and
K. D. Rothe, Int. Jour. Mod. Phys. {\bf A 14} (1999) 1163.

\bibitem{BR} L. V. Belvedere and K. D. Rothe, Mod. Phys. Lett. {\bf A 10}
(1995) 207.

%8
\bibitem{C} S. Coleman, Phys. Rev. {\bf D 11} (1975) 2088.

%9
\bibitem{Shaposnik} K. Furuya, R. E. Gamboa Sarav\'{\i} and F. A. Shaposnik, Nucl.
Phys. {\bf B}208 (1982) 159;

%10
\bibitem{Carena} M. Carena and C. E. M. Wagner, Int. Jour. Mod. Phys. {\bf A 6}
(1994) 253; K. Shizuya, Phys. Lett. {\bf B 213} (1988) 298; S. Miyake 
and K. Shyzuya, Phys. Rev. {\bf D 36} (1987) 3791; T. Berger, N. K. Falck and 
G. Kramer, Int. Jour. Mod. Phys. {\bf A 4} (1989) 427.

%11
\bibitem{Naon} C. M. Na\'on, Phys. Rev. {\bf D 31} (1985) 2035;

%12
\bibitem{Boy} D. Boyanovsky, I. Schmidt and M. F. L. Golterman, Annals of 
Physics {\bf 185},(1988) 111.

%13
\bibitem{Dutra} S. Coleman, R. Jackiw, L. Susskind, Ann. Phys. {\bf 93} (1975) 267; 
J. Fr\"ohlich, Phys. Rev. Lett. {\bf 34} (1975) 833; Comm. Math. Phys. {\bf 47} (1976) 233;
J. Fr\"ohlich, E. Seiler, Helv. Phys. Acta {\bf 49} (1976) 889;  I. Sachs and 
A. Wipf, Ann. Phys. {\bf 249} (1996) 380; A. S. Dutra, C. P. Natividade, H. Boschi-Filho, R. L.
P. G. Amaral and L. V. Belvedere, Phys. Rev. {\bf D 55} (1997) 49331.

%14
\bibitem{OS} K. Osterwalder, R. Schrader, Commun. Math. 
Phys. {\bf 31} (1973) 83; 
{\bf 42} (1975) 281; V. Glaser, Commun. Math. Phys. {\bf 37} (1974) 257.

%15
\bibitem{BA} L. V. Belvedere and R. L. P. G. Amaral, To appear in Phys. Rev. {\bf D} (2000).

%16
\bibitem{Fro} J. Fr\"ohlich, `` Non-Perturbative Quantum Field Theory'',
Advanced Series in Mathematical Physics, Vol 15, World Scientific, 1992.


%17
\bibitem{WZW} E. Witten, Comm. Math. Phys. {\bf 92} (1984) 455.

%18
\bibitem{PW} A. M. Polyakov and P. B. Wiegmann, Phys. Lett. {\bf B 131} (1983) 121;
Phys. Lett. {\bf B 141} (1984) 224.

%19
\bibitem{Mandelstam} S. Mandelstam, Phys. Rev. {\bf D 11} (1975) 3026.

%20
\bibitem{LS} S. Lowenstein and J. A. Swieca, Ann. Phys. {\bf 68}(1971)172.

%21
\bibitem{RS} K. D. Rothe and J. A. Swieca, Phys. Rev. {\bf D 15} (1977) 1675.

%22
\bibitem{S} J. A. Swieca, Forts. der Physik {\bf 25} (1977) 303.

%23
\bibitem{B} L. V. Belvedere, J. A. Swieca, K. D. Rothe and B. Schroer, Nucl. Phys.
{\bf B 153} (1979) 112; L. V. Belvedere,  Nucl. Phys.{\bf B 276} (1986) 197.

%24
\bibitem{AAR} E. Abdalla, M. C. Abdalla and K. D.
Rothe, ``Non-Perturbative Methods in $2$ Dimensional Quantum Field
Theory'', World Scientific Publishing, Singapure, 1991.

%25
\bibitem{WZ} J. Wess and B. Zumino, Phys. Lett. {\bf B 37} (1971) 95.

%7
\bibitem{JR}  R. Jackiw and R. Rajaraman, Phys. Rev. Lett. {\bf 54} (1985) 1219.


\end{thebibliography}
\end{document}